\documentclass[11pt]{article}

\usepackage[a4paper,
            bindingoffset=0.1in,
            left=0.8in,
            right=0.8in,
            top=0.8in,
            bottom=0.8in,
            footskip=.1in]{geometry}

\usepackage[dvipdfmx]{graphicx} 
\usepackage{bmpsize}
\usepackage[textwidth=8em,textsize=small]{todonotes}
\usepackage{amsmath}
\usepackage{natbib}
\usepackage{moreverb,url}
\usepackage[11pt]{moresize}
\usepackage{setspace}
\usepackage{threeparttable}
\usepackage{soul}
\usepackage{multirow}
\usepackage{amssymb}
\usepackage{xcolor}
\usepackage{mathtools}
\usepackage{algorithm}
\usepackage{algpseudocode}
\usepackage{caption}
\usepackage[absolute]{textpos}
\usepackage{booktabs}
\usepackage{wrapfig}
\usepackage{lscape}
\usepackage{rotating}
\usepackage{tablefootnote}
\usepackage{setspace} 
\usepackage{tablefootnote}
\usepackage{graphicx}
\usepackage{amsmath}
\usepackage{algorithm}
\usepackage{algpseudocode}
\usepackage{amsfonts}
\usepackage{bbm}
\usepackage{natbib}
\usepackage{booktabs}
\usepackage{multirow}
\usepackage{xcolor}

\usepackage[affil-it]{authblk}
\usepackage[english]{babel}
\usepackage{blindtext}

\def \bs { \boldsymbol }

\def \bX { \boldsymbol X }
\def \bZ { \boldsymbol Z }

\def \bB { \boldsymbol B }

\def \bD { \boldsymbol D }

\def \bS { \boldsymbol S }

\def \btheta { \boldsymbol \theta }

\def\bSig\mathbf{\Sigma}

\newcommand{\pkg}[1]{{\normalfont\fontseries{b}\selectfont #1}}
\let\proglang=\textsf
\let\code=\texttt

\begin{document}

\title{Propensity Score Analysis with Guaranteed Subgroup Balance}

\author[1]{Yan Li}
\author[2]{Yong-Fang Kuo}
\author[3,*]{Liang Li}

\affil[1]{Department of Quantitative Health Science, Division of Clinical Trials and Biostatistics, Mayo Clinic, Rochester, MN, USA }
\affil[2]{Department of Biostatistics and Data Science, The University of Texas Medical Branch, Galveston, TX, USA}
\affil[3]{Department of Biostatistics, The University of Texas MD Anderson Cancer Center, Houston, TX, USA}
\affil[*]{Corresponding author: Liang Li, LLi15@mdanderson.org}

\date{March, 2023}

\maketitle

\begin{abstract}
   Estimating the causal treatment effects by subgroups is important in observational studies when the treatment effect heterogeneity may be present. Existing propensity score methods rely on a correctly specified propensity score model. Model misspecification results in biased treatment effect estimation and covariate imbalance. We proposed a new algorithm, the propensity score analysis with guaranteed subgroup balance (G-SBPS), to achieve covariate mean balance in all subgroups. We further incorporated nonparametric kernel regression for the propensity scores and developed a kernelized G-SBPS (kG-SBPS) to improve the subgroup mean balance of covariate transformations in a rich functional class. This extension is more robust to propensity score model misspecification. Extensive numerical studies showed that G-SBPS and kG-SBPS improve both subgroup covariate balance and subgroup treatment effect estimation, compared to existing approaches. We applied G-SBPS and kG-SBPS to a dataset on right heart catheterization to estimate the subgroup average treatment effects on the hospital length of stay and a dataset on diabetes self-management training to estimate the subgroup average treatment effects for the treated on the hospitalization rate.
\end{abstract}

\section{Introduction}
\label{s:intro}

An important goal of observational studies is to estimate the treatment effect. Naive comparison between treatment groups is subject to selection bias when covariates are unbalanced between treatment groups due to lack of randomization. Propensity score, the conditional probability of treatment assignment given the covariates, is widely used to adjust for covariate imbalance and remove selection bias through matching \citep{stuart2010matching}, stratification \citep{rosenbaum1984reducing}, regression \citep{vansteelandt2014regression}, and weighting \citep{lunceford2004stratification}. Commonly used estimands of the treatment effect include the average treatment effect (ATE) and the average treatment effect for the treated (ATT). When there are heterogeneous treatment effects (HTEs), subgroups with different characteristics respond to the treatment differently. For example, a drug may have better efficacy on patients with certain genetic traits. The overall treatment effects that ignore the underlying heterogeneity, such as the ATE or ATT, do not provide sufficient granular information for scientific investigation and clinical practice. The HTE is common in biomedical, epidemiological, and social research. In this paper, we study HTEs among pre-specified subgroups of scientific interest, and these subgroups are defined through covariates.

The subgroup HTEs can be estimated with or without modeling the outcome. Examples of the former approach include regression models stratified on subgroups, Bayesian additive regression trees \citep{hill2011bayesian, chipman2010bart}, causal forest \citep{wager2018estimation}, etc. However, their performance depends on a correctly specified outcome model. Additionally, there are benefits to being blinded from the outcome data when developing causal models \citep{Rubin2008}. In this paper, we focus on the latter approach and study the causal subgroup analysis, an HTE estimation method that adjusts subgroup covariate imbalance in a propensity score analysis \citep{dong2020subgroup, yang2021propensity}. Our proposed approach is built upon propensity score weighting.

The propensity score is a balancing score, i.e., the treatment assignments and covariates are independent conditional on the propensity score \citep{rosenbaum1983central}. Theoretically, the propensity score balances covariates in the overall population and any covariate-defined subgroups. We define \textbf{overall balance} as the mean difference of covariates or their transformations between the two treatments in the overall population. This is the type of covariate balance that most published propensity score methods deal with. We define \textbf{subgroup balance} as the mean difference of covariates or their transformations between the two treatments in subgroups. This is the focus of this paper. The propensity scores are unknown and must be estimated from a parametric or nonparametric model, such as logistic regression, covariate balancing propensity score (CBPS; \citep{imai2014covariate}), boosting (GBM; \citep{mccaffrey2004propensity}), or covariate balancing scoring rules (CBSR; \citep{zhao2019covariate}). When the estimated model deviates from the true propensity score model or has an explicit or hidden lack of fit, the estimated propensity scores do not have the desired balancing property \citep{lipropensity}. This scenario may result in a lack of overall or subgroup balance. The latter is usually more severe because the majority of propensity score analysis procedures are developed to achieve overall balance. The subgroups have smaller sample sizes and hence are more prone to covariate imbalance, and there are often many subgroups under simultaneous consideration. For example, when the estimated propensity model is misspecified, the CBPS ensures exact overall balance by method design but subgroup covariate imbalance may still arise, which causes bias in the subgroup treatment effect estimation \citep{dong2020subgroup}. 

Nonparametric propensity score models, such as boosting \citep{mccaffrey2004propensity}, random forest \citep{lee2010improving} and CBSR \citep{zhao2019covariate}, do not guarantee overall and subgroup balance \citep{dong2020subgroup, yang2021propensity, li2021propensity, lipropensity}. Although the nonparametric methods may reduce model misspecification and bias due to the higher flexibility than their parametric counterparts, the estimation may have more variability, a typical bias-variance trade-off phenomenon. This has been observed, for example, in the comparison between CBSR and CBPS \citep{li2021propensity}. This trade-off is amplified in the subgroup analysis due to the large number of subgroups under research and the limited sample size of each subgroup. 

Developing methods that ensure both overall and subgroup covariate balance is essential when studying subgroup HTEs. For this purpose, \cite{dong2020subgroup} proposed the subgroup balancing propensity score (SBPS). SBPS selects among either parametric logistic regression models with covariate-by-subgroup interactions fitted to the overall sample or parametric logistic regression models fitted to the subgroup samples. However, this method cannot guarantee subgroup balance when the propensity score model is misspecified or when the sampling variability results in extreme inverse probability weights. The SBPS needs to examine up to $2^R$ logistic regression models with various combinations of covariates and covariate-by-subgroup interactions, where $R$ is the number of subgroups. When $R$ is large, the computational burden can be tremendous. Furthermore, the SBPS requires mutually exclusive subgroups, which limits its use in statistical practice. \cite{yang2021propensity} extended the overlap weights, originally developed for exact overall balance \citep{LiFan2017}, to the subgroup analysis. The overlap weights require a deviation from the widely used estimands such as ATE and ATT. When the propensity score model is misspecified, the overlap weights produce biased estimation despite showing no signs of covariate imbalance \citep{Mao2018}. 

In this paper, we propose the propensity score weighting analysis with guaranteed subgroup balance (G-SBPS), which optimizes both overall and subgroup balance simultaneously. The G-SBPS does not require mutually exclusive subgroups. We estimate the propensity scores by solving a system of equations that achieves the mean independence between the treatment indicator and the covariate terms in the propensity score model, which includes covariates, subgroup indicators and their interactions. We show that the G-SBPS guarantees both overall and subgroup balance. To further improve the flexibility of propensity score models and reduce misspecification, we extend the G-SBPS to nonparametric estimation by using kernel principle component analysis (PCA). We propose a parameter tuning algorithm tailored for the subgroup analysis, which optimizes the subgroup covariate balance while controlling the overall balance. This kernelized G-SBPS (kG-SBPS) optimizes the overall and subgroup balance of the covariates and their transformations from a rich functional class. In simulations and two empirical data applications, both the G-SBPS and kG-SBPS demonstrated robustness to model misspecification compared to existing propensity score methods or subgroup propensity score analysis methods.  

The rest of the paper is organized as follows. Section 2 presents the model, the G-SBPS algorithm, the kG-SBPS algorithm, and the tuning algorithm. Section 3 evaluates the numerical performance of G-SBPS and kG-SBPS and compares them with other published methods in a simulation study. Section 4 presents two real data applications, one for the estimation of ATE and one for ATT. We conclude this paper with a summary and discussion in Section 5. Some tables and figures are included in the online supplementary materials and numbered as Table S1, Fig. S1, etc.

\section{Methodology}
\subsection{Model set-up}
We consider a sample of $N$ observations with $N_0$ untreated subjects, denoted by $T_i=0$, and $N_1$ treated subjects, denoted by $T_i=1$. For each subject, we observe a vector of $M$ covariates $\bZ_i = (Z_{i1}, \dots, Z_{iM})^T$ and the outcome variable $Y_i$. The observed outcome is $Y_i = T_i Y_{i}(1) + (1-T_i)Y_{i}(0)$, where $Y_{i}(1)$ and $Y_{i}(0)$ are two potential outcomes corresponding to the treated and untreated, respectively. Let the pre-specified subgroups of interest be denoted by $\bS_i = (S_{i1}, \dots, S_{iK})^T$, where $S_{ik} = 1$ if the $i^{th}$ subject belongs to the $k^{th}$ ($k = 1,\dots,K$) subgroup and $0$ otherwise. The subgroup variables $\{ S_{ik} \}$ are functions of the covariates $\bZ_i$. For example, $S_{ik} = I(Z_{i1} > 50 ~\text{years}, Z_{i2} = \text{male}, Z_{i3} = \text{no diabetes})$ represents a male subject who is older than 50 and has no diabetes. The $K$ groups do not need to be mutually exclusive, and each subject can belong to multiple subgroups, i.e. $\sum_{k=1}^K S_{ik} \neq 1$. Let $N_k$ be the number of subjects in the $k^{th}$ subgroup with $N_{0k}$ untreated and $N_{1k}$ treated. 

The propensity score of subject $i$ is denoted by $p(\bZ_i) = P(T_i = 1 \mid \bZ_i)$. We use standard assumptions for propensity score analysis \citep{rosenbaum1983central}, including the stable unit treatment value assumption (SUTVA), the no unmeasured confounding assumption, and the overlap assumption ($0 < P(\bZ) <1$). The propensity score is a balancing score, i.e. the treatment assignment is independent of the covariates conditional on the propensity score: $ T \perp \bZ \mid p(\bZ) $. The treatment assignment can be viewed as being ``randomized'' within a small neighborhood of the propensity scores. Thus, the propensity score can balance covariates in not only the overall sample but also the subgroups. The overall and subgroup treatment effects, such as ATE or ATT, can be estimated without bias. In the next two subsections, we present the parametric G-SBPS model for the propensity scores, and the nonparametric kG-SBPS model. The presentation focuses on how these models guarantee covariate balance in the subgroups.

\subsection{Parametric propensity score model with guaranteed subgroup balance (G-SBPS)}

This subsection discusses G-SBPS as a parametric approach. We model the propensity score by the logit link function as
\begin{equation} \label{eq6}
    \pi_i = p_{\btheta}(\bZ_i, \bS_i) = p_{\btheta}(\bX_i) = \frac{\exp(\bX_i^T\btheta )}{1+\exp(\bX_i^T\btheta)}, \notag
\end{equation}
where $\bX_i = \phi (\bZ_i, \bS_i)=[1, \bS_i^T, \bZ_i^T, S_{i1}\bZ_i^T, \dots, S_{iK}\bZ_i^T]$ is the design vector that includes the observed covariates $\bZ_i$, the subgroup indicator $\bS_i$ and their interactions. 

We discuss ATE estimation first. To optimize the overall and subgroup balance, we consider the following loss functions proposed in CBSR \citep{zhao2019covariate, li2021propensity}.
\begin{equation} \label{eq7}
\begin{split}
L_{ATE} & = \sum^{N}_{i=1} T_i \left[\log \left( \dfrac{p_{\btheta}(\bX_i)}{1-p_{\btheta}(\bX_i)} \right)-\dfrac{1}{p_{\btheta}(\bX_i)} \right] \\
 & + \sum^{N}_{i=1} (1-T_i)\left[ \log \left( \dfrac{1-p_{\btheta}(\bX_i)}{p_{\btheta}(\bX_i)} \right) - \dfrac{1}{1-p_{\btheta}(\bX_i)} \right] 
\end{split}
\end{equation}

The propensity scores are estimated by maximizing this loss function with respect to the modeling parameters $\btheta$, and the estimated parameters are consistent \citep{zhao2019covariate}. When $\pi_i=p_{\btheta}(\bX_i)$ is twice continuously differentiable with respect to $\btheta$, the corresponding estimating equation is
\begin{equation} \label{eq9}
    \dfrac{\partial L_{ATE}}{\partial \btheta} = \sum^{N}_{i=1}B_{ATE}(\btheta|T_i,\bX_i) =  \sum^{N}_{i=1} \left( \dfrac{T_i}{p_{\btheta}(\bX_i)}-\dfrac{1-T_i}{1-p_{\btheta}(\bX_i)} \right)\bX_i = 0 .
\end{equation} 
Notably, Equation (\ref{eq9}) is the overall covariate balancing condition, as studied in CBPS \citep{imai2014covariate}. However, our design vector $\bX_i$ includes an intercept, the covariates $\bZ_i$, the subgroup indicator $\bS_i$, and the interactions between the covariates and subgroup indicators $\bZ_i \times \bS_i$. The balancing conditions corresponding to $\bZ_i$ ensure the overall balance, and more importantly, the balancing conditions corresponding to $\bZ_i \times \bS_i$ guarantee the subgroup balance. The conditions corresponding to the intercept and $\bS_i$ ensure an equal total sum of weights between the treated and untreated in the overall population and subgroups. There are a total of $(1+K)(1+M)$ of balance conditions and parameters.

The loss function in (\ref{eq7}) is concave with a global maximum. The Hessian matrix ( the matrix of second-order partial and cross-partial derivatives) is:
\begin{equation} \label{eq11}
    H_{ATE}(\btheta) = \sum_{i=1}^{N} -[T_i\exp(-\bX_i^T \btheta)+(1-T_i)\exp(\bX_i^T \btheta)] \bX_i \bX_i^T, \notag
\end{equation} 
which is negative semi-definite. Estimating $\btheta$ by solving the balance conditions in (\ref{eq9}) is equivalent to maximizing equation (\ref{eq7}), which produces globally optimal solutions. 

Therefore, we choose to estimate the coefficient $\btheta$ in the propensity score model by solving the balance conditions using the generalized method of moments (GMM), which was utilized by CBPS \citep{imai2014covariate}. Specifically, we use the following GMM estimators,
\begin{equation} \label{eq13}
    \hat{\btheta}_{\text{GMM}} =  \underset{\btheta \in \Theta}{\mathrm{argmin}}~ \bar{B}(\btheta|T, \bX)^T\Sigma(T,\bX)^{-1}\bar{B}(\btheta|T, \bX)
\end{equation}
\begin{equation} \label{eq14}
    \bar{B}(\btheta|T, \bX) =  \displaystyle{\frac{1}{N}\sum_{i=1}^n} B_{ATE}(\btheta |T_i,\bX_i) \notag ,
\end{equation}
where $\bX_i = \phi(\bZ_i, \bS_i)$ includes the covariates, subgroup indicators, and their interactions. We choose a consistent covariance estimator of $\bar{B}(\btheta|T, \bX)$, which is given by
\begin{equation} \label{eq16}
    \Sigma(T,\bX) = \bX \bX^T .
\end{equation}
We used the ``continuous updating'' GMM estimator, which has been shown to have better finite sample properties than ``two-step optimal'' GMM estimator \citep{hansen1996finite}. 

Next, we discuss the estimation procedure for the ATT. The procedure is similar to the ATE but with the following changes. The CBSR loss function, the estimating equation, and the Hessian matrix become
\begin{equation} \label{eq8}
\begin{aligned}
    L_{ATT} = & ~ \sum^{N}_{i=1}\{T_i\log(\frac{p_{\btheta}(\bX_i)}{1-p_{\btheta}(\bX_i)}) - \frac{1-T_i}{1-p_{\btheta}(\bX_i)} \}, \\
    \frac{\partial L_{ATT}}{\partial \btheta} = & ~ \sum^{N}_{i=1}B_{ATT}(\btheta|T_i,\bX_i) =  \sum^{N}_{i=1}(T_i-(1-T_i)\frac{p_{\btheta}(\bX_i)}{1-p_{\btheta}(\bX_i)})\bX_i = 0, \\
    H_{ATT}(\btheta) = & ~ \sum_{i=1}^{N} -(1-T_i)\exp(\bX_i^T \btheta) \bX_i \bX_i^T . \notag
\end{aligned}
\end{equation}
Because $H_{ATT}(\btheta)$ are negative semi-definite, $\btheta$ can be estimated as a global maximum of $L_{ATT}$, which is also the solutions to the balance equations $\dfrac{\partial L_{ATT}}{\partial \btheta} = 0$. 

The proposed G-SBPS avoids some of the limitations of SBPS mentioned in the Introduction section. The parameter estimation of G-SBPS produces global optimization, which guarantees the minimally possible overall and subgroup imbalance. In data analysis practice, we often observed that the G-SBPS produced exact subgroup balance. The SBPS stochastically searches through a large number of parametric propensity score models to find one with the best overall and subgroup balance among the candidate models. This process does not produce globally optimal solution or guaranteed exact global or subgroup balance.

\subsection{Nonparametric propensity score model with guaranteed subgroup balance (kG-SBPS)} 

A misspecified propensity score model leads to covariate imbalance and bias in estimated treatment effects \citep{li2021propensity}. A commonly used practice to alleviate this problem is to tweak the logistic model by adding covariate transformations or interaction terms, until a satisfactory covariate balance is achieved. However, this process is \textit{ad hoc}, lacks methodologically justified guidelines, and does not guarantee good balance. Some methods that force overall covariate balance (e.g., \citep{imai2014covariate}) are subject to model misspecification and biased treatment effect estimation \citep{Mao2018}.  

We propose to improve the flexibility of the propensity score model in G-SBPS through reproducing kernel Hilbert space (RKHS), which transforms the observed covariate vector into an $N$-dimensional vector of features on which overall and subgroup balance can be achieved. Specifically, we use the kernel PCA (\citealp{scholkopf2002learning}). First, we construct the kernel matrix $\bs K_{N \times N}$ by producing a measure of similarity between any two subjects using a pre-specified kernel function. In this paper, we use the Gaussian kernel $k(\bZ_i, \bZ_j) = \exp(-\lVert \bZ_i-\bZ_j \rVert ^2/\sigma)$. Second, we conduct eigen-decomposition of the kernel matrix, i.e., $\bs K = \bs P \bD \bs P^T$. The feature space containing the covariate transformations is $\bs \omega(\bZ) = \bs P \bD^{1/2}$. Each column of $\bs \omega(\bZ)$ represents one transformed feature of the covariate matrix $\bZ$. Because we would like to balance the most informative features, we only select a finite number of columns in $\bs \omega(\bZ)$, corresponding to $99\%$ variance calculated from the eigenvalues. We denote the transformed feature space as $\bs \omega_{l_{99}}(\bZ)$. We replace the observed covariates $\bZ_i$ in equation (\ref{eq13}) of G-SBPS with the new transformed features $\bs \omega_{l_{99}}(\bZ)$ such that $\bX_i = [1,\bS_i^T,\bs \omega_{l_{99}}(\bZ)_i^T, S_{i1}\bs \omega_{l_{99}}(\bZ)_i^T, \dots, S_{iK}\bs \omega_{l_{99}}(\bZ)_i^T]$. We aim to achieve overall and subgroup balances in those transformed covariates instead of the original covariates. The bandwidth of the Gaussian kernel $\sigma$, which determines the set of covariate transformations, is tuned by the algorithm below. 

The existing subgroup propensity score methods, such as the SBPS \citep{dong2020subgroup}, models the propensity score parametrically. Although it produces improved subgroup balance, it suffers from propensity score model misspecification (shown in the simulation results below). Nonparametric methods not designed for subgroup analysis, such as boosting \citep{mccaffrey2004propensity}, often give unsatisfactory subgroup balance, leading to biased subgroup treatment effect estimation \citep{dong2020subgroup, yang2021propensity}. The kernelized G-SBPS is the first subgroup propensity score analysis method that aims at both goals: flexible nonparametric modeling and guaranteed overall and subgroup balance. 

When modeling the propensity score without using the outcome, the commonly used out-of-sample target for optimization, such as prediction error, does not always produce overall or subgroup balance, which may lead to suboptimal performance in treatment effect estimation. We propose to optimize the subgroup balance, while controlling the overall balance. The standardized difference (S/D) is used to measure covariate balance  \citep{griffin2017chasing, cannas2019comparison}. The S/D is the absolute difference in weighted mean between treated and untreated groups, divided by the pooled standard deviation of the weighted data \citep{li2013weighting}. In addition to the S/D of covariates in the overall population, we optimize the S/D in the subgroups. The details of this tuning process are presented in the supplementary materials (Algorithm 1). We use this algorithm to tune the bandwidth $\sigma$ of the kernelized G-SBPS. Specifically, we set the range of $\sigma$ to be the $0.1$ and $0.9$ quantile of the Euclidean distances between samples. The 20 candidate values of $\sigma$ are equally spaced on the log scale within this range. We choose the $\sigma$ that optimizes subgroup balance while controlling the overall balance. We name the analytical framework that couples the kernelized G-SBPS and this tuning process as \textbf{kG-SBPS}.

\section{Simulations}

In this section, we compare the proposed G-SBPS (parametric method) and kG-SBPS (nonparametric method) with two popular propensity score methods and two representative subgroup propensity score analysis methods under various numerical settings.

\subsection{Simulation design}

Let $G$ be the subgroup indicator, taking values in $\{ 1, 2,..., K \}$. We use $K = 4$. The sample size of the $k$-th subgroup is $N_k = 500$. There are four covariates. $X_1 \sim N(\mu_k, 1)$, where $\mu_k = 3-3(k-1)/(K-1)$. $X_2 \sim \mathrm{Uniform}(0,1)$. $X_3 \sim N(0,1)$. $X_4 \sim \mathrm{Bernoulli}(0.4)$. We consider two propensity score model specifications:  
\begin{itemize}
\item[PS1:] (Correct PS model) The propensity score model is a logistic regression with the main effects of covariates and subgroup-specific intercept: 
$$
\text{logit}(\pi) = \sum_{k=1}^K \delta_k \mathbbm{1} \{G = k\} + \beta_1 X_1 + \beta_2 X_2 +\beta_3 X_3 +\beta_4 X_4
$$
where $\bs\beta = ( -0.2, -0.2, 0.4, -0.4 )$.  
\item[PS2:] (Misspecified PS model) The propensity score model has additional interaction and nonlinear terms: 
$$
\text{logit}(\pi) = \sum_{k=1}^K \delta_k \mathbbm{1} \{G = k\} + \beta_1 X_1 + \beta_2 X_2 +\beta_3 X_3 +\beta_4 X_4 + \beta_5 X_1^2 + \beta_6 X_1X_4
$$
where $\bs \beta = (-1.5,-0.5,0.5,-0.5,0.5,0.5)$ for ATE estimation and\\ $\bs \beta = (-1.5,-0.8,0.2,-0.8,0.5,0.5)$ for ATT estimation. 
\end{itemize}
In practice, the data analyst usually just uses the main effect terms. With many covariates, there are numerous possible interactions and nonlinear terms, and it is difficult to determine which ones should be added to the model. In this simulation study, we call PS1 the ``correct PS model'' because our parametric data analysis uses this model, and call PS2 the ``misspecified PS model'' because the parametric analysis does not include any interaction or nonlinear terms. For both PS1 and PS2, $\delta_k = -1+2(k-1)/(K-1)$. 

We consider two outcome models:
\begin{itemize}
\item[OM1:] (Standard outcome model): The outcome model includes the main effects of all covariates in the propensity score model. The treatment effects $\{ \eta_k \}$ vary across subgroups.    
$$
Y = 200 + \sum^{K}_{k=1}\eta_k[\mathbbm{1}(G=k) T] + 20X_1 + 10X_2 + 10X_3 + 10X_4+\epsilon
$$
\item[OM2:] (Extended outcome model): The outcome model includes additional interactions and nonlinear transformations of the covariates. These additional terms are unknown to the data analyst and hence are not explicitly accounted for in the data analysis.
 $$
Y = 200 + \sum^{K}_{k=1}\eta_k[\mathbbm{1}(G=k) T] + 20X_1 + 10X_2 + 10X_3 + 10X_4-5X_1^2 + 10X_1X_4 + \epsilon
$$
\end{itemize}
The residual is $\epsilon \sim N(0,1)$. The true subgroup treatment effects are $\eta_k = -10+20(k-1)/(K-1)$. As stated above, our parametric propensity score model includes all the main effects of covariates. The theory of \cite{hazlett2018kernel} suggests that consistent treatment effect estimation can be achieved even when the propensity score model is misspecified, as long as the propensity score adjustment balances all the linear terms in the outcome model. Therefore, the OM1 and OM2 help us observe a performance difference in the different methods under model misspecification. 

The two propensity score models (PS1, PS2) and the two outcome models (OM1, OM2) produce four scenarios. We simulated data under each scenario, and evaluated the performance of the following six methods in the subgroup treatment effect estimation and overall and subgroup balance. 
\begin{itemize}
    \item[(a)] Logistic: the logistic regression analysis with the main effects of observed covariates and the subgroup indicator (\proglang{R} package \pkg{glm}).
    \item[(b)] Logistic-S: separately fitted logistic models within each subgroup. Each model includes the main effects of covariates. This was studied in \cite{dong2020subgroup} and implemented in \proglang{R} package \pkg{WeightIt}.
    \item[(c)] CBPS: the just-identified CBPS with the main effects of covariates and the subgroup indicator (\proglang{R} package \pkg{CBPS}).
    \item[(d)] SBPS: the SBPS approach of \cite{dong2020subgroup}, implemented in the \code{SBPS} function of \proglang{R} package \pkg{WeightIt} with the main effects of covariates.
    \item[(e)] G-SBPS: the proposed parametric G-SBPS method with the main effects of covariates.
    \item[(f)] kG-SBPS: the proposed nonparametric kG-SBPS method.
\end{itemize}

We studied both ATE and ATT estimation these are widely used estimands. The treatment effect estimators were evaluated by percent bias and root mean squared error (RMSE) in each subgroup. The overall or subgroup covariate balance were quantified by the S/D of covariates in the overall population or the subgroups. In each simulation scenario, the results were aggregated from 500 Monte Carlo repetitions.

\subsection{Covariate balance}

First, we examine the overall balance. When the propensity score model was correctly specified (PS1), all methods under comparison had good overall balance in the sense that the S/Ds were generally less than $5\%$ (Fig S1 and S2). This result held when the estimand was either ATE or ATT. CBPS and G-SBPS had nearly zero imbalance in $X_1$-$X_4$ because they both used the correct model and had balance constraints on these covariates. The balance is slightly worse with the interaction terms because they are not in the propensity score model used by logistic, logistic-S, CBPS, SBPS, and G-SBPS. The kG-SBPS, in contrast, achieved notably better balance in these two interaction terms because it is a nonparametric method and it has built-in balance constraints on a large number of covariate transformations. The SBPS and G-SBPS are both parametric methods, but the latter had better performance, probably due to its more effective balance control. While G-SBPS had constraints on subgroup balance and CBPS had constraints on overall balance, the G-SBPS achieved comparable overall balance as the CBPS, because good subgroup balance implies good overall balance. 

When the propensity score model was misspecified (PS2), the only method that maintained good performance is the kG-SBPS because it is the only nonparametric method. CBPS and G-SBPS maintained good balance in $X_1$-$X_4$, because this was what their balance constraints were designed for. However, the enlarged imbalance in the interaction terms contradicted with the property of the propensity scores, suggesting that the ``twisting'' of a parametric propensity score model to satisfy the balance constraints under misspecification may inadvertently create scores that are not propensity scores. Therefore, checking the goodness-of-fit is as important as checking the balance when using a parametric model for propensity score. 

Next, we study the subgroup balance and subgroup treatment effects. Fig 1-2 present the results from ATE estimation, and we comment on those results here. The results from ATT estimation are presented in Fig S3-S4 and they produce similar conclusions. Comparing Fig 1 and Fig 2, we observe the expected results that all parametric methods performed better under the correct PS model (PS1). The subgroup S/Ds are usually higher than the corresponding overall S/Ds because the subgroups have smaller sample sizes. Nonetheless, the subgroup S/Ds are generally less than 10\% under the PS1. The Logistic-S and SBPS methods performed better than Logistic and CBPS in subgroup balance, because the former methods were designed for subgroup propensity score analysis. This is the opposite of the overall balance results in Fig S1, where the latter methods were better. The proposed G-SBPS and kG-SBPS had equivalent or better performance than all other methods in terms of subgroup balance.  

Under the misspecified PS model (PS2), Logistic, Logistic-S, CBPS and SBPS have deteriorated subgroup balance performance in all covariates and their transformations, as expected. The G-SBPS still gives exact subgroup balance of $X_1$-$X_4$ despite the propensity score model misspecification, because it enforces the subgroup balance constraints. This is a step forward in the protection against model misspecification compared to the other 4 methods. The G-SBPS does not properly balance $X_1^2$ and $X_1 X_4$ terms because they are not in the propensity score model. This deficiency was addressed by the kG-SBPS, which produced the best subgroup balance results in all scenarios.  

In addition, to evaluate the effect of small sample size on the performance of the proposed methods, we applied all methods on the simulations with a correct PS model (PS 1) with only 40 units for subgroup 2. We found that only G-SBPS achieves good subgroup balance for subgroup 2 if $5\%$ of S/D is used as the threshold; this was observed in both the ATE and ATT estimation (Figure S7 and S8). This finding is supported by the theoretical justification that the estimation of G-SBPS is globally optimal. Therefore, G-SBPS always leads to the smallest weighted mean difference between two treatments in the covariate-defined subpopulations. Intriguingly, kG-SBPS does not achieve subgroup balance in this case, which suggests that it needs a larger sample size for accurate estimation, as expected from a nonparametric method.

\subsection{Treatment effect estimation}

Tables 1 and 2 show the estimation of subgroup ATEs under the correct or misspecified PS models. Lower \% bias and smaller RMSE indicate better performance.  The two methods that explicitly deal with subgroup balance (Logistic-S, SBPS) perform better than the ones that do not (Logistic, CBPS). This is consistent with the subgroup covariate balance results above. Compared with the other four methods (Logistic, Logistic-S, CBPS, SBPS), the G-SBPS has the best \% bias and RMSE, due to its use of the globally optimal solution to the subgroup balance constraints (Section 2). The kG-SBPS has more variation than the G-SBPS, due to its nonparametric nature. Nonetheless, the kG-SBPS still has better \% bias and RMSE than the other four methods. 

The benefit of kG-SBPS is best shown when the model is under misspecification, where it is the only method that performs well in all scenarios. Even under misspecification, the G-SBPS did not break down in all scenarios like the other four methods: it still performed well under standard outcome model. This is due to the theory in \cite{hazlett2018kernel}, which states that even when a propensity score model is misspecified, as long as it balances the linear additive terms in the outcome model (which is the case here with the standard outcome model), the misspecification does not cause bias to the treatment effect estimation. Of the three parametric subgroup propensity score analysis methods (Logistic-S, SBPS, G-SBPS), only our proposed method exploited this theoretical result, which produces the doubly-robust-like performance shown in Table 2. This is because the optimal solution from the G-SBPS results in exact subgroup balance, while Logistic-S and SBPS do not have this guarantee. 

The estimation of ATT is presented in the online supplementary materials (Table S1 and S2). The results are similar, demonstrating better performance of the proposed G-SBPS and kG-SBPS methods over the four other existing methods (Logistic, Logistic-S, SBPS, CBPS). The G-SBPS has no bias under the standard outcome model but some bias under the extended outcome model. The kG-SBPS shows some bias in ATT estimation under both outcome models, although the bias is generally smaller than the other four existing methods. Increasing the subgroup sample size to 1000 per group can reduce the bias of kG-SBPS to no or slight bias in ATT estimation, which suggests that kG-SBPS requires a larger sample size as a nonparametric method (Table S3).

\section{Data Applications}
\subsection{Right Heart Catheterization (RHC) data}

We applied the proposed methods, G-SBPS and kG-SBPS, to the right heart catheterization (RHC) data \citep{connors1996effectiveness} to examine the average treatment effect (ATE) of RHC vs. non-RHC on the length of hospital stay. The data set contains $5,735$ subjects, including $2184$ receiving RHC and $3551$ who did not (non-RHC). The study design was reported previously \citep{connors1996effectiveness, hirano2001estimation}. We excluded one subject from the analysis due to a missing outcome value. The observed covariates include demographic characteristics, comorbidity conditions, lab test results, etc. Among the 72 covariates, 57 were tested to have statistically significant mean difference between the RHC and non-RHC groups, and these covariates were included in our subgroup analysis.

For illustration, we studied the subgroup treatment effects of RHC in a non-overlapping subgroup scheme and an overlapping subgroup scheme (Table S4). In the former scheme, three non-overlapping subgroups (``3-subgroup scheme'') were defined from mean blood pressure ($<$ 80, 80-120,or  $>$ 120mmHg); each patient belongs to only one subgroup.  In the latter scheme, six overlapping subgroups were defined (``6-subgroup scheme''), with three based on the mean blood pressure and another three based on the estimated probability of surviving 2 months. These six subgroups overlap because a patient can belong to a blood pressure subgroup and a survival probability subgroup simultaneously. Here the estimated 2 months survival probability was calculated using the SUPPORT prognostic model \citep{knaus1995support}. It is known that high blood pressure can induce cardiovascular damage, which may lead to worse prognosis after RHC treatment. In addition, patients with lower estimated survival probability are usually sicker and may need longer hospital stay after the RHC treatment. These are the motivations to study those subgroups. The Logistic-S and SBPS methods do not work with overlapping subgroups. Therefore, we compared all six methods in the simulation among the three non-overlapping subgroups, but excluded Logistic-S and SBPS from the analysis with the overlapping subgroups.  

All methods achieved overall covariate balance for the two subgroup schemes in the sense that the S/Ds were generally less than 5\%, with CBPS and G-SBPS being the best (Fig. S5). The subgroup balance results are in Figure 3. For the 3-subgroup scheme in Fig 3(a), all methods give reasonably good subgroup balance. The two subgroup analysis methods, Logistic-S and SBPS, performed better than Logistic and CBPS, but performed worse than the proposed methods, GSBPS and k-GSBPS. For the 6-subgroup scheme in Fig 3(b), G-SBPS and kG-SBPS had notably better subgroup balance than Logistic and CBPS. On average, G-SBPS has slightly better subgroup balances than kG-SBPS, which may be attributed to the larger variation in kG-SBPS results, a typical bias-variance trade-off phenomenon between parametric vs. nonparametric methods. These observations are consistent with the results of simulation studies. 

Before propensity score adjustment, the average lengths of hospital stay in the RHC was on average 4.20, 6.28, 8.22, 3.54, 5.59 and 4.87 days longer than in the non-RHC in the subgroups 1 to 6, respectively. The estimated subgroup ATEs by various methods are reported in Table S5. They are considerable differences in the estimated ATEs across subgroups, but these results are ignored in the usual propensity score analysis. This example highlights the importance of exploring subgroup treatment effects. The estimated subgroup ATEs are generally smaller after the propensity score adjustment, regardless which method was used. Notably, the estimated ATEs by the four subgroup analysis methods are smaller than the the general propensity score methods (Logistic and CBPS), which may suggest that the improved subgroup covariate balance reduced heterogeneity between the RHC and non-RHC and hence also reduced bias. The ATEs of subgroup 1 (low blood pressure, $<80$) and subgroup 2 (normal blood pressure, $80-120$) are similar after propensity score adjustment, but they are smaller than the ATE of subgroup 3 (high blood pressure, $> 120$). These results show that the RHC causes longer hospital stay among patients with higher blood pressure. The RHC causes longer hospital stay in subgroup with highest mortality risk (subgroup4), whose baseline health may be worse. 



\subsection{Diabetes Outpatient Self-management Training Services (DSMT) data}

We applied G-SBPS and kG-SBPS to a second dataset to illustrate their performance in estimating the ATT. The ATT is best applied to situations where the treated group has a much smaller sample size than the control group. We used Texas Cancer Registry-Medicare linkage data for Diabetes Self-management Training (DSMT) program among 5$+$ year cancer survivors with diabetes, aged 66$+$, and alive in 2006-2019 \citep{lee2023utilization}. If patients were eligible for multiple years, the first year meeting the eligibility criteria was selected as the index year. The treatment variable is the receipt of the DSMT training vs. not. The outcome variable is the hospitalization rate within 3 years of the index year. The original data contain $3,348$ patients who received DSMT for the first time and $71,307$ controls. We excluded around $2\%$ patients whose race is not identified as Hispanic, White or Black. The resulting dataset includes $3,283$ treated patients (DSMT) and $69,871$ controls (no DSMT) in 2006-2018. As a pre-processing step to balance key patient characteristics, we performed 1 to 5 matches between treated and untreated subjects by gender, the type of diabetes, and incident diabetes. The resulting matched data contains all $3,283$ treated units and $16,401$ matched controls. Table S6 presents the summary statistics of the baseline covariates in the matched data. Baseline demographic covariates were extracted from Medicare enrollment file in the index year. Baseline comorbidities and diabetic complications were identified from inpatient and outpatient claims in the year before the index year. This study examines the effect of DSMT training on the hospitalization rate. 

We are interested in the ATTs in subgroups of various social statuses, which are described in Table S7. Patients eligible for dual Medicaid or living in the metropolitan area have easier access to medical care. But we also speculate that patients living in rural areas or city may have different lifestyles, which can contribute to the heterogeneity in the effect of DSMT training. Similarly, married and unmarried patients might respond to DSMT training differently due to the differences in their way of life or personalities. There is overlap among the six subgroups. Again, Logistic-S and SBPS were only applied to the non-overlapped subgroups 1-4 (Table S7). All subgroups have relatively sufficient sample sizes except subgroup 4, which allows us to study the proposed method with a small subgroup size. 

For both overlapped and non-overlapped subgroups, global balances are attained by all methods, with less than $1\%$ of S/D in most covariates (Fig S6). However, the subgroup balances are not achieved by all methods (Fig 4). For subgroup 1, all methods gave good balance. For subgroups 2 to 4, Logistic and CBPS did not achieve good balance for all covariates at the benchmark level of $10\%$ S/D. Logistic-S, SBPS, G-SBPS, and kG-SBPS achieved good balance, although G-SBPS and kG-SBPS resulted in even smaller S/D on average. For the overlapping subgroups 5 and 6, G-SBPS and kG-SBPS achieved better balance than other methods. However, the balances of Logistic and CBPS are reasonably well. In summary, Logistic and CBPS produced the worst subgroup balance, while G-SBPS and kG-SBPS resulted in the best subgroup balance. 

Before propensity score adjustment, the average rates of 3-year hospitalization from the index year are $5.9\%$, $18.7\%$, $6.5\%$, $33.7\%$, $11.7\%$ and $2.7\%$ lower in the DSMT than the control, from subgroups 1 to 6 respectively. Generally, the effect size becomes smaller after propensity score adjustment (Table S8). The impact of DSMT is most pronounced among individuals residing in metropolitan areas and possessing a dual-eligible health plan during the index year, and least among the population characterized by both factors being negative (Table S7 and S8), following propensity score adjustment. These observations support our hypothesis that easier access to medical care may boost the effect of DSMT. For subgroup 1 with the largest sample size, the estimated ATTs are around $4\%$ after adjustment of all methods. For subgroups 2, 3, and 4, the estimated ATTs are similar for the four subgroup analysis methods, but larger for the general propensity score methods (Table S8). We attribute this result to the better subgroup covariate balance from the subgroup analysis methods (Table S6). DSMT training has a larger effect on the rate of hospitalization among the unmarried subgroups (Table S8). The ATTs were reduced to around $3\%$ by G-SBPS and kG-SBPS and $1\%$ by Logistic and CBPS among unmarried subgroups (Table S8). 





\section{Discussion}

Subgroup causal effect estimation has wide application, but received limited attention in the propensity score analysis field. Previously, we demonstrated that global covariate balance is not equivalent to having propensity score's balancing property when the fitted propensity score model is subject to misspecification \cite{lipropensity}. Thus, propensity score methods that optimize the global balance, such as CBPS, may result in subgroup imbalance and biased subgroup treatment effect estimation. There are critical limitations in the current subgroup analysis methods. Firstly, subgroup analysis methods, such as SBPS, are not applicable to overlapped subgroups. Secondly, while the SBPS can improve the subgroup balance (as shown in the numerical studies in this paper), it suffers from suboptimal parameter estimation and may not ensure adequate subgroup balance. We propose the novel G-SBPS (parametric method) and kG-SBPS (nonparametric method) that guarantee exact subgroup balance through globally optimal parameter estimations. Our numerical studies demonstrated that the proposed methods significantly outperform the existing methods in the literature. 

G-SBPS shows a doubly-robust-like property, i.e., if the fitted propensity model coincides with either the true propensity score model or the true outcome model, the estimated treatment effect shows no bias and small RMSE. Our simulations demonstrate this (Tables 1, 2, S1 and S2). Being a nonparametric method, kG-SBPS requires large subgroup sample sizes for its good performance (Table S3, Fig S7 and S8). However, kG-SBPS is more robust to model misspecification, especially when both the propensity score and outcome models are unknown (Table 2, S2 and S3). Hence, it is necessary to conduct model diagnostics, particularly when the subgroup sample sizes are small. If the data analysts are confident that the fitted model agrees with either the true propensity score or outcome model, G-SBPS should perform the best because it has less variability than the kG-SBPS. 

In future work, it is necessary to develop new model selection methods for the subgroup propensity score analysis. It is also important to study the sensitivity of G-SBPS and kG-SBPS on the level of model misspecification, effect sizes, and the subgroup sample sizes. Last, there are various potential extensions beyond the scope of this paper. For example, G-SBPS and kG-SBPS can be extended for subgroup analysis with multiple treatment groups.

\section*{Acknowledgements}

This research is supported by NIH grants R01CA225646 and P30CA016672, and CPRIT grant RP210130. \vspace*{-8pt}

\label{lastpage}

\bibliographystyle{unsrtnat}
\bibliography{SubgroupMain}

\begin{thebibliography}{29}
\providecommand{\natexlab}[1]{#1}
\providecommand{\url}[1]{\texttt{#1}}
\expandafter\ifx\csname urlstyle\endcsname\relax
  \providecommand{\doi}[1]{doi: #1}\else
  \providecommand{\doi}{doi: \begingroup \urlstyle{rm}\Url}\fi

\bibitem[Stuart(2010)]{stuart2010matching}
Elizabeth~A. Stuart.
\newblock Matching methods for causal inference: A review and a look forward.
\newblock \emph{Statistical Science}, 25\penalty0 (1):\penalty0 1--21, February
  2010.

\bibitem[Rosenbaum and Rubin(1984)]{rosenbaum1984reducing}
Paul~R Rosenbaum and Donald~B Rubin.
\newblock Reducing bias in observational studies using subclassification on the
  propensity score.
\newblock \emph{Journal of the American Statistical Association}, 79\penalty0
  (387):\penalty0 516--524, September 1984.

\bibitem[Vansteelandt and Daniel(2014)]{vansteelandt2014regression}
Stijn Vansteelandt and Rhian~M Daniel.
\newblock On regression adjustment for the propensity score.
\newblock \emph{Statistics in Medicine}, 33\penalty0 (23):\penalty0 4053--4072,
  May 2014.

\bibitem[Lunceford and Davidian(2004)]{lunceford2004stratification}
Jared~K Lunceford and Marie Davidian.
\newblock Stratification and weighting via the propensity score in estimation
  of causal treatment effects: a comparative study.
\newblock \emph{Statistics in Medicine}, 23\penalty0 (19):\penalty0 2937--2960,
  August 2004.

\bibitem[Hill(2011)]{hill2011bayesian}
Jennifer~L Hill.
\newblock Bayesian nonparametric modeling for causal inference.
\newblock \emph{Journal of Computational and Graphical Statistics}, 20\penalty0
  (1):\penalty0 217--240, 2011.

\bibitem[Chipman et~al.(2010)Chipman, George, and McCulloch]{chipman2010bart}
Hugh~A Chipman, Edward~I George, and Robert~E McCulloch.
\newblock Bart: Bayesian additive regression trees.
\newblock \emph{The Annals of Applied Statistics}, 4\penalty0 (1):\penalty0
  266--298, 2010.

\bibitem[Wager and Athey(2018)]{wager2018estimation}
Stefan Wager and Susan Athey.
\newblock Estimation and inference of heterogeneous treatment effects using
  random forests.
\newblock \emph{Journal of the American Statistical Association}, 113\penalty0
  (523):\penalty0 1228--1242, 2018.

\bibitem[Rubin(2008)]{Rubin2008}
Donald~B. Rubin.
\newblock For objective causal inference, design trumps analysis.
\newblock \emph{The Annals of Applied Statistics}, 2\penalty0 (3), September
  2008.
\newblock \doi{10.1214/08-aoas187}.
\newblock URL \url{https://doi.org/10.1214/08-aoas187}.

\bibitem[Dong et~al.(2020)Dong, Zhang, Zeng, and Li]{dong2020subgroup}
Jing Dong, Junni~L Zhang, Shuxi Zeng, and Fan Li.
\newblock Subgroup balancing propensity score.
\newblock \emph{Statistical methods in medical research}, 29\penalty0
  (3):\penalty0 659--676, 2020.

\bibitem[Yang et~al.(2021)Yang, Lorenzi, Papadogeorgou, Wojdyla, Li, and
  Thomas]{yang2021propensity}
Siyun Yang, Elizabeth Lorenzi, Georgia Papadogeorgou, Daniel~M Wojdyla, Fan Li,
  and Laine~E Thomas.
\newblock Propensity score weighting for causal subgroup analysis.
\newblock \emph{Statistics in Medicine}, 40\penalty0 (19):\penalty0 4294--4309,
  2021.

\bibitem[Rosenbaum and Rubin(1983)]{rosenbaum1983central}
Paul~R Rosenbaum and Donald~B Rubin.
\newblock The central role of the propensity score in observational studies for
  causal effects.
\newblock \emph{Biometrika}, 70\penalty0 (1):\penalty0 41--55, 1983.

\bibitem[Imai and Ratkovic(2014)]{imai2014covariate}
Kosuke Imai and Marc Ratkovic.
\newblock Covariate balancing propensity score.
\newblock \emph{Journal of the Royal Statistical Society: Series B (Statistical
  Methodology)}, 76\penalty0 (1):\penalty0 243--263, 2014.

\bibitem[McCaffrey et~al.(2004)McCaffrey, Ridgeway, and
  Morral]{mccaffrey2004propensity}
Daniel~F. McCaffrey, Greg Ridgeway, and Andrew~R. Morral.
\newblock Propensity score estimation with boosted regression for evaluating
  causal effects in observational studies.
\newblock \emph{Psychological Methods}, 9\penalty0 (4):\penalty0 403--425,
  December 2004.

\bibitem[Zhao(2019)]{zhao2019covariate}
Qingyuan Zhao.
\newblock Covariate balancing propensity score by tailored loss functions.
\newblock \emph{The Annals of Statistics}, 47\penalty0 (2):\penalty0 965--993,
  2019.

\bibitem[Li and Li(2023)]{lipropensity}
Yan Li and Liang Li.
\newblock Propensity score analysis with local balance.
\newblock \emph{Statistics in Medicine}, 42\penalty0 (15):\penalty0 2637--2660,
  April 2023.
\newblock \doi{10.1002/sim.9741}.
\newblock URL \url{https://doi.org/10.1002/sim.9741}.

\bibitem[Lee et~al.(2009)Lee, Lessler, and Stuart]{lee2010improving}
Brian~K. Lee, Justin Lessler, and Elizabeth~A. Stuart.
\newblock Improving propensity score weighting using machine learning.
\newblock \emph{Statistics in Medicine}, 29\penalty0 (3):\penalty0 337--346,
  December 2009.

\bibitem[Li and Li(2021)]{li2021propensity}
Yan Li and Liang Li.
\newblock Propensity score analysis methods with balancing constraints: A monte
  carlo study.
\newblock \emph{Statistical Methods in Medical Research}, 30\penalty0
  (4):\penalty0 1119--1142, 2021.

\bibitem[Li et~al.(2017)Li, Morgan, and Zaslavsky]{LiFan2017}
Fan Li, Kari~Lock Morgan, and Alan~M. Zaslavsky.
\newblock Balancing covariates via propensity score weighting.
\newblock \emph{Journal of the American Statistical Association}, 113\penalty0
  (521):\penalty0 390--400, November 2017.
\newblock \doi{10.1080/01621459.2016.1260466}.
\newblock URL \url{https://doi.org/10.1080/01621459.2016.1260466}.

\bibitem[Mao et~al.(2018)Mao, Li, and Greene]{Mao2018}
Huzhang Mao, Liang Li, and Tom Greene.
\newblock Propensity score weighting analysis and treatment effect discovery.
\newblock \emph{Statistical Methods in Medical Research}, 28\penalty0
  (8):\penalty0 2439--2454, June 2018.
\newblock \doi{10.1177/0962280218781171}.
\newblock URL \url{https://doi.org/10.1177/0962280218781171}.

\bibitem[Hansen et~al.(1996)Hansen, Heaton, and Yaron]{hansen1996finite}
Lars~Peter Hansen, John Heaton, and Amir Yaron.
\newblock Finite-sample properties of some alternative {GMM} estimators.
\newblock \emph{Journal of Business {\&} Economic Statistics}, 14\penalty0
  (3):\penalty0 262--280, 1996.

\bibitem[Sch{\"o}lkopf and Smola(2002)]{scholkopf2002learning}
Bernhard Sch{\"o}lkopf and Alexander~J Smola.
\newblock \emph{Learning with kernels: support vector machines, regularization,
  optimization, and beyond}.
\newblock The MIT Press, 2002.

\bibitem[Griffin et~al.(2017)Griffin, McCaffrey, Almirall, Burgette, and
  Setodji]{griffin2017chasing}
Beth~Ann Griffin, Daniel~F McCaffrey, Daniel Almirall, Lane~F Burgette, and
  Claude~Messan Setodji.
\newblock Chasing balance and other recommendations for improving nonparametric
  propensity score models.
\newblock \emph{Journal of Causal Inference}, 5\penalty0 (2), 2017.

\bibitem[Cannas and Arpino(2019)]{cannas2019comparison}
Massimo Cannas and Bruno Arpino.
\newblock A comparison of machine learning algorithms and covariate balance
  measures for propensity score matching and weighting.
\newblock \emph{Biometrical Journal}, 61\penalty0 (4):\penalty0 1049--1072,
  2019.

\bibitem[Li and Greene(2013)]{li2013weighting}
Liang Li and Tom Greene.
\newblock A weighting analogue to pair matching in propensity score analysis.
\newblock \emph{The International Journal of Biostatistics}, 9\penalty0
  (2):\penalty0 215--234, 2013.

\bibitem[Hazlett(2020)]{hazlett2018kernel}
Chad Hazlett.
\newblock Kernel balancing: A flexible non-parametric weighting procedure for
  estimating causal effects.
\newblock \emph{Statistica Sinica}, 30:\penalty0 1155--1189, 2020.

\bibitem[Connors et~al.(1996)Connors, Speroff, Dawson, Thomas, Harrell, Wagner,
  Desbiens, Goldman, Wu, Califf, Fulkerson, Vidaillet, Broste, Bellamy, Lynn,
  and Knaus]{connors1996effectiveness}
Alfred F~Jr Connors, Theodore Speroff, Neal~V Dawson, Charles Thomas, Frank
  E~Jr Harrell, Douglas Wagner, Norman Desbiens, Lee Goldman, Albert~W Wu,
  Robert~M Califf, William J~Jr Fulkerson, Humberto Vidaillet, Steven Broste,
  Paul Bellamy, Joanne Lynn, and William~A Knaus.
\newblock The effectiveness of right heart catheterization in the initial care
  of critically ill patients. {SUPPORT} investigators.
\newblock \emph{{JAMA}: The Journal of the American Medical Association},
  276\penalty0 (11):\penalty0 889--897, 1996.

\bibitem[Hirano and Imbens(2001)]{hirano2001estimation}
Keisuke Hirano and Guido~W Imbens.
\newblock Estimation of causal effects using propensity score weighting: an
  application to data on right heart catheterization.
\newblock \emph{Health Services and Outcomes Research Methodology}, 2\penalty0
  (3-4):\penalty0 259--278, 2001.

\bibitem[Knaus et~al.(1995)Knaus, Harrell, Lynn, et~al.]{knaus1995support}
WA~Knaus, FE~Harrell, J~Lynn, et~al.
\newblock The support prognostic model: Prediction of survival for seriously
  ill hospitalized patients.
\newblock \emph{Ann Intern Med}, 122:\penalty0 191--203, 1995.

\bibitem[Lee et~al.(2023)Lee, Digbeu, Serag, Sallam, and
  Kuo]{lee2023utilization}
Wei-Chen Lee, Biai Dominique~Elmir Digbeu, Hani Serag, Hanaa Sallam, and
  Yong-Fang Kuo.
\newblock Utilization of diabetes self-management program among breast,
  prostate, and colorectal cancer survivors: Using 2006--2019 texas medicare
  data.
\newblock \emph{Plos One}, 18\penalty0 (7):\penalty0 e0289268, 2023.

\end{thebibliography}

\clearpage

\begin{center}
	\begin{table}[t]%
            \footnotesize
		\centering
		\caption{The performance of various methods in the estimation of \textbf{subgroup ATEs} in the simulation. The \textbf{correct PS model (PS1)} was used. The true treatment effect for subgroup 1 to 4 are $-10$, $-10/3$, $10/3$ and $10$, respectively. The results were aggregated from 500 Monte Carlo repetitions. \label{tab1}}%
		\begin{tabular*}{450pt}{@{\extracolsep\fill}ccccccccc@{\extracolsep\fill}}
			\toprule
			 & \textbf{Outcome}  & \textbf{Subgroup}  & \textbf{Logistic}  & \textbf{Logistic-S} & \textbf{CBPS} & \textbf{SBPS} & \textbf{G-SBPS} & \textbf{kG-SBPS} \\
			\midrule
			\multirow{8}{*}{\%Bias}& \multirow{4}{*}{Standard}& \textbf{1} & -0.35 & 0.23 & -0.04 & 0.23 & -0.09 & -1.45 \\&& \textbf{2} & -4.85 & -1.56 & -4.54 & -1.56 & -0.17 & -1.20 \\ && \textbf{3} & -1.41 & 0.37 & -1.53 & 0.37 & 0.22 & 0.59 \\ && \textbf{4} & -0.80 & -0.31 & -0.91 & -0.31 & -0.01 & 0.38 \\ 
            \cline{2-9}
            &\multirow{4}{*}{Extended}& \textbf{1} & -1.76 & -1.49 & -1.72 & -1.49 & -1.45 & -1.23 \\&& \textbf{2} & -3.95 & -2.33 & -3.60 & -2.33 & -2.07 & -2.43 \\ && \textbf{3} & 0.23 & 3.13 & 0.06 & 3.13 & 2.95 & 0.76 \\ && \textbf{4} & -1.15 & -0.71 & -1.21 & -0.71 & -0.38 & 0.14 \\ 
			\midrule
			\multirow{8}{*}{RMSE}& \multirow{4}{*}{Standard}& \textbf{1} & 2.98 & 1.34 & 2.59 & 1.34 & 0.14 & 0.42 \\&& \textbf{2} & 1.97 & 0.67 & 2.01 & 0.67 & 0.11 & 0.25 \\ && \textbf{3} & 1.92 & 0.26 & 2.11 & 0.26 & 0.10 & 0.18 \\ && \textbf{4} & 2.03 & 0.47 & 2.13 & 0.47 & 0.10 & 0.20 \\ 
            \cline{2-9}
            &\multirow{4}{*}{Extended}& \textbf{1} & 3.48 & 2.02 & 3.10 & 2.02 & 1.31 & 0.60 \\&& \textbf{2} & 1.98 & 1.13 & 1.98 & 1.13 & 0.96 & 0.27 \\ && \textbf{3} & 1.87 & 0.85 & 2.02 & 0.85 & 0.82 & 0.23 \\ && \textbf{4} & 2.53 & 1.10 & 2.62 & 1.10 & 0.88 & 0.33 \\ 	
			\bottomrule
		\end{tabular*}
	\end{table}
\end{center}

\begin{center}
	\begin{table}[t]%
            \footnotesize
		\centering
		\caption{The performance of various methods in the estimation of \textbf{subgroup ATEs} in the simulation. The \textbf{misspecified PS model (PS2)} was used. The true treatment effect for subgroup 1 to 4 are $-10$, $-10/3$, $10/3$ and $10$, respectively. The results were aggregated from 500 Monte Carlo repetitions. \label{tab2}}%
		\begin{tabular*}{450pt}{@{\extracolsep\fill}ccccccccc@{\extracolsep\fill}}
			\toprule
			 & \textbf{Outcome}  & \textbf{Subgroup}  & \textbf{Logistic}  & \textbf{Logistic-S} & \textbf{CBPS} & \textbf{SBPS} & \textbf{G-SBPS} & \textbf{kG-SBPS} \\
			\midrule
			\multirow{8}{*}{\%Bias}& \multirow{4}{*}{Standard}& \textbf{1} & -204.95 & 108.82 & -207.29 & 32.86 & -0.14 & -5.20 \\&& \textbf{2} & -272.65 & 197.57 & -286.45 & 180.88 & -0.15 & -7.14 \\ && \textbf{3} & -291.67 & 32.14 & -277.67 & 32.14 & -0.11 & -0.61 \\ && \textbf{4} & -193.97 & -17.49 & -191.34 & -17.49 & -0.15 & -2.64 \\ 
            \cline{2-9}
            &\multirow{4}{*}{Extended}& \textbf{1} & 32.62 & 53.51 & 33.52 & 37.22 & 33.16 & 2.34 \\&& \textbf{2} & -9.94 & 216.10 & -14.31 & 205.92 & 114.39 & 1.19 \\ && \textbf{3} & 317.42 & -73.53 & -305.91 & -73.53 & -95.91 & -4.97 \\ && \textbf{4} & -236.97 & -47.23 & -234.09 & -47.23 & -29.74 & -6.52 \\ 
			\midrule
			\multirow{8}{*}{RMSE}& \multirow{4}{*}{Standard}& \textbf{1} & 20.57 & 16.21 & 20.80 & 12.73 & 0.15 & 0.66 \\&& \textbf{2} & 9.45 & 7.56 & 9.85 & 7.44 & 0.11 & 0.38 \\ && \textbf{3} & 9.95 & 1.30 & 9.49 & 1.30 & 0.09 & 0.21 \\ && \textbf{4} & 19.49 & 2.20 & 19.23 & 2.20 & 0.12 & 0.41 \\ 
            \cline{2-9}
            &\multirow{4}{*}{Extended}& \textbf{1} & 3.94 & 8.61 & 4.03 & 4.93 & 3.91 & 0.46 \\&& \textbf{2} & 2.16 & 8.02 & 2.12 & 7.57 & 4.05 & 0.29 \\ && \textbf{3} & 10.78 & 2.65 & 10.40 & 2.65 & 3.33 & 0.29 \\ && \textbf{4} & 23.80 & 5.09 & 23.51 & 5.09 & 3.24 & 0.79 \\ 	
			\bottomrule
		\end{tabular*}
	\end{table}
\end{center}

\begin{figure}
\centerline{\includegraphics[scale = 1]{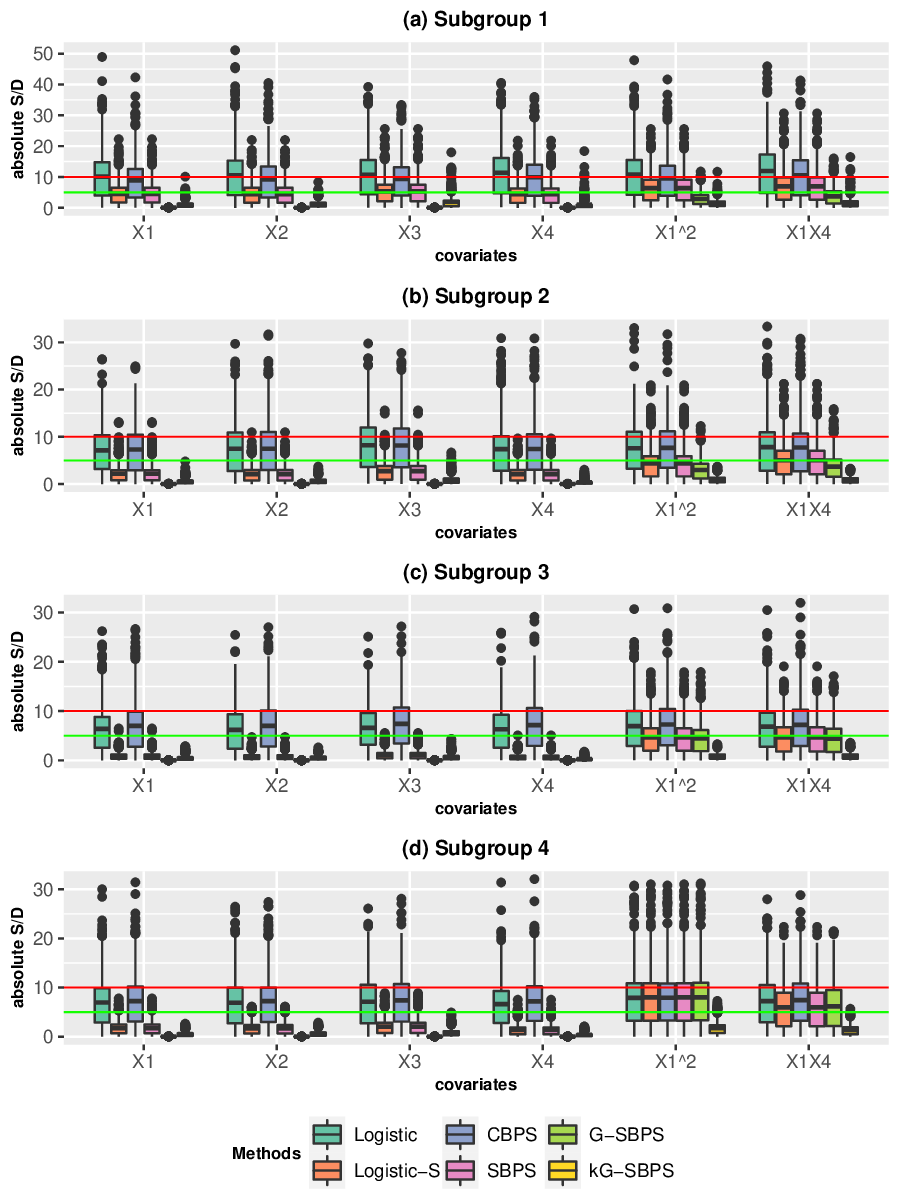}}
 \caption{Boxplots of the standardized differences (S/D; \%) in the four subgroups when estimating the \textbf{ATE} in the simulation studies. The data are simulated from the \textbf{Correct PS model} (PS1). The boxplots show the distribution of S/D from 500 Monte Carlo repetitions. Red line: 10\% S/D; Green line: 5\% S/D. }
\end{figure}

\begin{figure}
\centerline{\includegraphics[scale = 1]{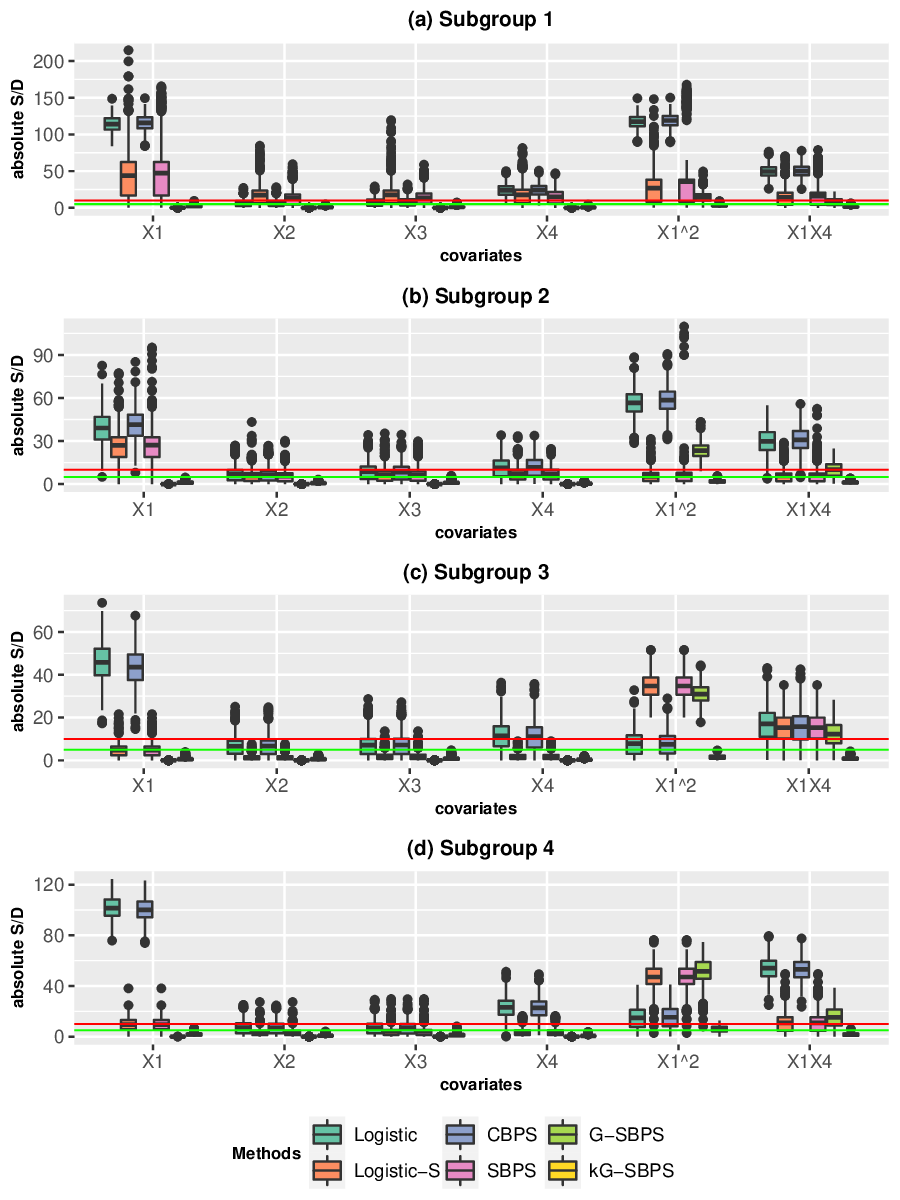}}
 \caption{Boxplots of the standardized differences (S/D; \%) in the four subgroups when estimating the \textbf{ATE} in the simulation studies. The data are simulated from the \textbf{Misspecified PS model} (PS2). The boxplots show the distribution of S/D from 500 Monte Carlo repetitions. Red line: 10\% S/D; Green line: 5\% S/D. }
\end{figure}

\begin{figure}
\centerline{\includegraphics[scale = 0.9]{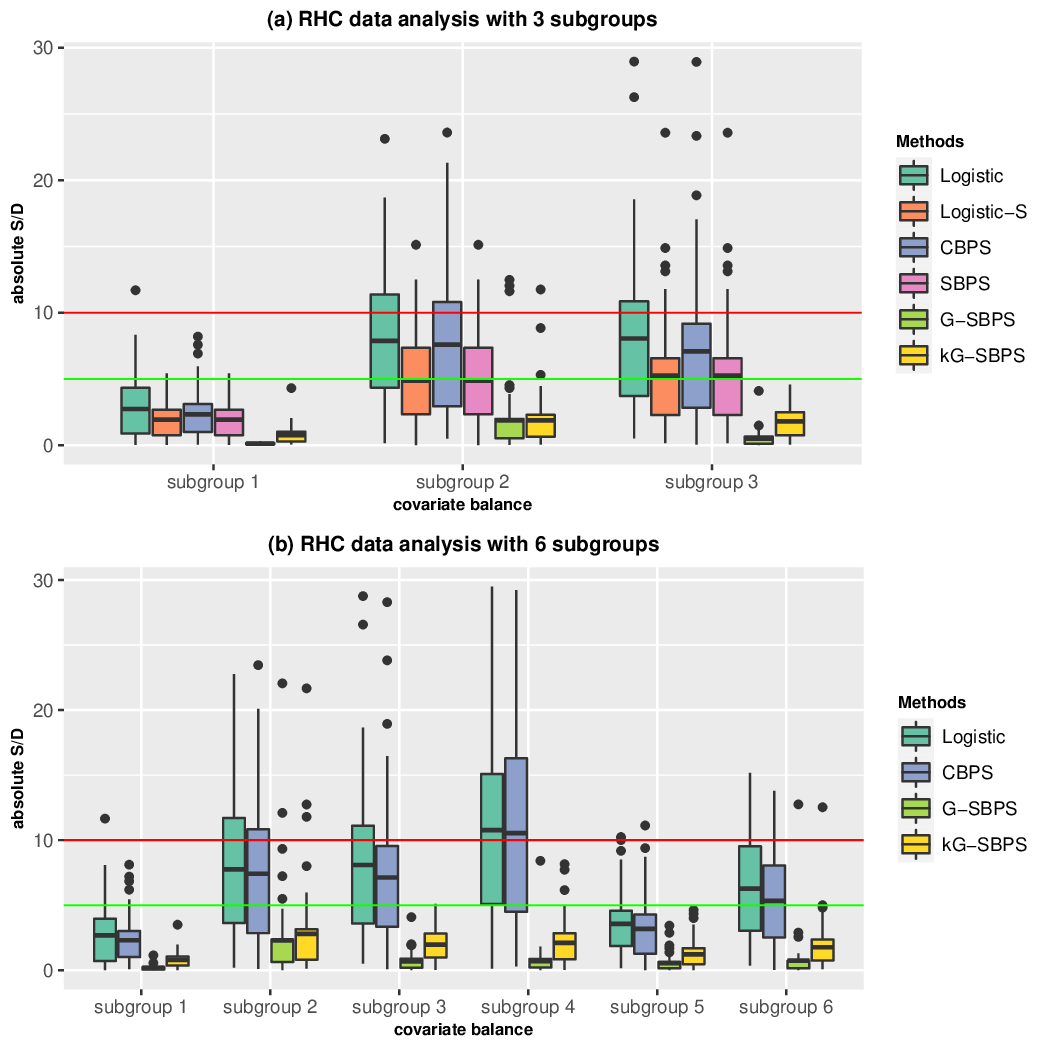}}
 \caption{Boxplots of the subgroup standardized differences (S/D) of all covariates in the RHC data analysis. Red line: 10\% S/D; Green line: 5\% S/D. }
\end{figure}

\begin{figure}
\centerline{\includegraphics[scale = 0.9]{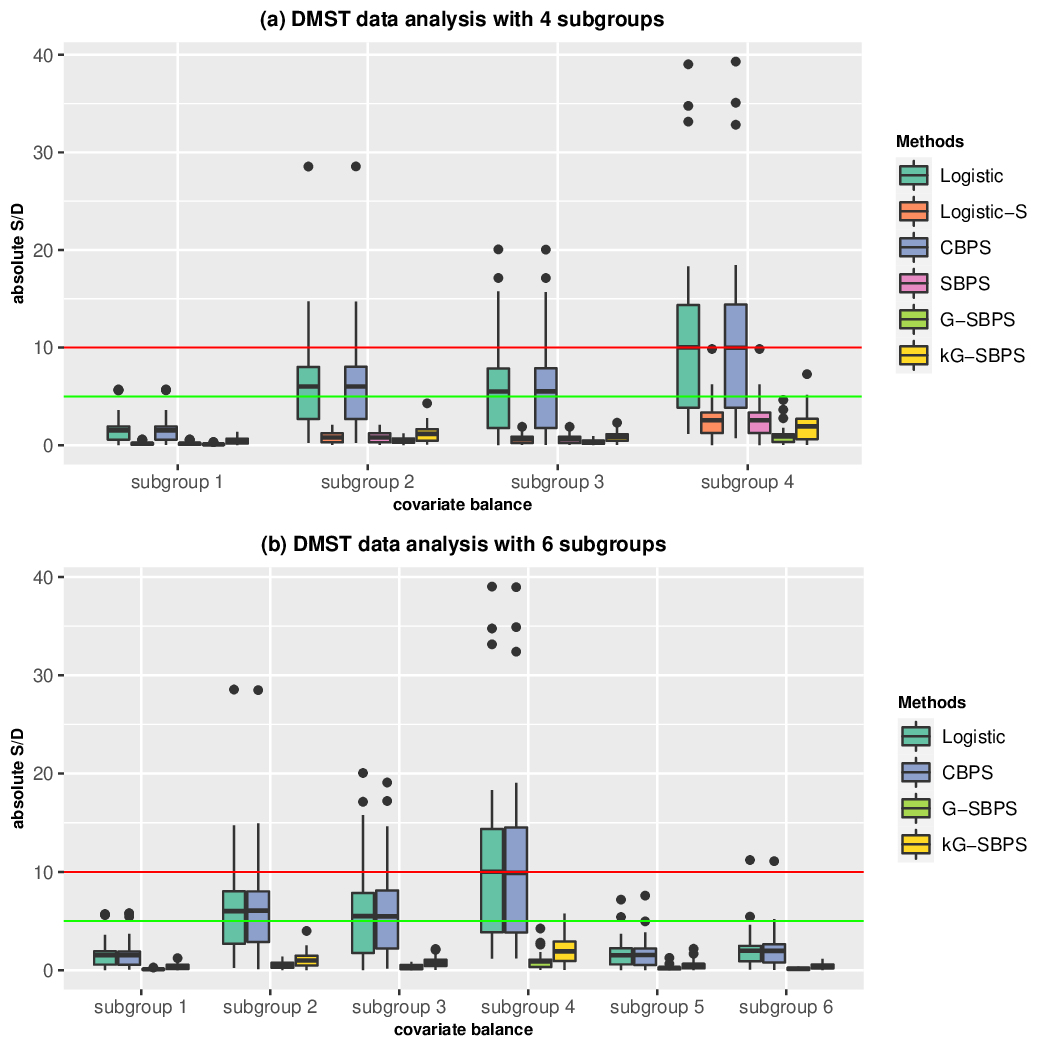}}
 \caption{Boxplots of the subgroup standardized differences (S/D) of all covariates in the DSMT data analysis. Red line: 10\% S/D; Green line: 5\% S/D.}
\end{figure}

\clearpage

\section{Supplemental Tables and Figures}

\begin{algorithm}
\caption{\textbf{}: Choosing $\sigma$ in kG-SBPS}\label{euclid}
\begin{algorithmic}[1]
\State For each $\sigma \in \{\sigma_1, \dots, \sigma_{20} \}$.
\State Compute IPW: $\{ \hat{p}_{\btheta}(\bs \omega_{l_{99}}(\bZ)_i,\bS_i); i=1,2,...,N \} $ using the kG-SBPS in Section 2.3 and 2.4.
\State Compute $w_{i1} = 1/\hat{p}_{\btheta}(\bs \omega_{l_{99}}(\bZ)_i,\bS_i)$ and $w_{i0} = 1/(1-\hat{p}_{\btheta}(\bs \omega_{l_{99}}(\bZ)_i,\bS_i))$ for ATE estimation or $w_{i1} = 1$ and $w_{i0} = \hat{p}_{\btheta}(\bs \omega_{l_{99}}(\bZ)_i,\bS_i)/(1-\hat{p}_{\btheta}(\bs \omega_{l_{99}}(\bZ)_i,\bS_i))$ for ATT estimation.
\State Evaluate global balance: calculate the S/D of each observed covariates $\bZ$ in the overall population such that
$$
\bs B_{\text{global}} = 100 \times \left|  \frac{\sum_{i=1}^N w_{i1}\bZ_i T_i}{\sum_{i=1}^N w_{i1}T_i}-\frac{\sum_{i=1}^N w_{i0}\bZ_i(1-T_i)}{\sum_{i=1}^N w_{i0}(1-T_i)} \right| /sd(\bZ),
$$
where $sd(\bZ)$ are the pooled standard deviations of $\bZ$ between two treatment groups such that $sd(\bZ) = \sqrt{(sd_{(1)}^2+sd_{(0)}^2)/2}$, where $sd_{(1)}$ and $sd_{(0)}$ are the weighted standard deviations of $\bZ$
$$
sd_{(1)}^2 = \frac{\sum w_{i1} T_i}{(\sum w_{i1} T_i)^2-\sum w_{i1}^2 T_i }\sum w_{i1}T_i\left(\bZ_i - \frac{w_{i1} \bZ_i T_i}{\sum w_{i1} Ti}\right)^2
$$
$$
sd_{(0)}^2 = \frac{\sum w_{i0} (1-T_i)}{(\sum w_{i0} (1-T_i))^2-\sum w_{i0}^2 (1-T_i) }\sum w_{i0}(1-T_i)\left(\bZ_i - \frac{w_{i0} \bZ_i (1-T_i)}{\sum w_{i0}(1- Ti)}\right)^2.
$$
\State Evaluate subgroup balance: calculate the S/D of the observed covariates in the covariate-defined subpopulations such that 
$$
\bB_{S_k} = 100 \times \left| \frac{\sum w_{i1}\bZ_i T_i}{\sum w_{i1} T_i} -\frac{\sum w_{i0}\bZ_i (1-T_i)}{\sum w_{i0}(1- T_i)} \right| / sd_k(\bZ) \bigm|_{S_{ik}=1},
$$
where  $S_k = \{ i | S_{ik}=1 \}$. And $sd_k(\bZ)$ is calculated from subjects in each subpopulation.
\State Output $l^* = \text{arg min}$ $\frac{1}{MK}\displaystyle{\sum_{m=1}^{M}}\displaystyle{\sum_{k=1}^{K}} \bB_{S_k}$ subject to $\text{max} (\bs B_{\text{global}}) \leq 10\%$. Output ``fail'' if $\text{max} (\bs B_{\text{global}}) > 10\%$.
\end{algorithmic}
\end{algorithm}

\clearpage

\setcounter{table}{0} 
\begin{center}
	\begin{table}[h]%
            \renewcommand{\thetable}{S\arabic{table}}
            \footnotesize
		\centering
		\caption{The performance of various methods to estimate \textbf{subgroup ATTs} in the simulations with the \textbf{correct PS model (PS1, see Section 3.1)}. The true treatment effect for subgroup 1 to 4 are $-10$, $-10/3$, $10/3$, and $10$, respectively. The number of Monte Carlo repetitions is 500. \label{tabs1}}%
		\begin{tabular*}{450pt}{@{\extracolsep\fill}ccccccccc@{\extracolsep\fill}}
			\toprule
			 & \textbf{Outcome}  & \textbf{Subgroup}  & \textbf{Logistic}  & \textbf{Logistic-S} & \textbf{CBPS} & \textbf{SBPS} & \textbf{G-SBPS} & \textbf{kG-SBPS} \\
			\midrule
			\multirow{8}{*}{\%Bias}& \multirow{4}{*}{Standard}& \textbf{1} & -0.23 & -0.12 & -0.59 & -0.12 & -0.10 & 0.22 \\&& \textbf{2} & -3.88 & 0.57 & -4.93 & 0.57 & -0.18 & -0.25 \\ && \textbf{3} & -1.29 & 0.15 & 0.01 & 0.15 & 0.23 & 0.59 \\ && \textbf{4} & -0.99 & -0.54 & -0.70 & -0.54 & -0.02 & 0.67 \\ 
            \cline{2-9}
            &\multirow{4}{*}{Extended}& \textbf{1} & -1.61 & -0.95 & -1.28 & -0.95 & -1.09 & -0.62 \\&& \textbf{2} & -3.38 & -0.57 & -3.42 & -0.57 & -0.75 & -0.35 \\ && \textbf{3} & 0.24 & 2.46 & 1.23 & 2.46 & 2.40 & -0.07 \\ && \textbf{4} & -1.37 & -0.98 & -1.18 & -0.98 & -0.39 & 0.15 \\ 
			\midrule
			\multirow{8}{*}{RMSE}& \multirow{4}{*}{Standard}& \textbf{1} & 2.78 & 0.34 & 3.08 & 0.34 & 0.14 & 0.24 \\&& \textbf{2} & 1.99 & 0.42 & 2.22 & 0.42 & 0.11 & 0.17 \\ && \textbf{3} & 2.01 & 0.58 & 2.09 & 0.58 & 0.10 & 0.16 \\ && \textbf{4} & 2.15 & 0.84 & 1.67 & 0.84 & 0.10 & 0.25 \\ 
            \cline{2-9}
            &\multirow{4}{*}{Extended}& \textbf{1} & 2.82 & 1.11 & 3.02 & 1.11 & 1.12 & 0.34 \\&& \textbf{2} & 1.84 & 0.98 & 2.02 & 0.98 & 0.94 & 0.25 \\ && \textbf{3} & 1.98 & 1.01 & 2.02 & 1.01 & 0.82 & 0.25 \\ && \textbf{4} & 2.70 & 1.44 & 2.18 & 1.44 & 0.91 & 0.42 \\ 	
			\bottomrule
		\end{tabular*}
	\end{table}
\end{center}
\clearpage

\begin{center}
	\begin{table}[t]%
            \renewcommand{\thetable}{S\arabic{table}}
            \footnotesize
		\centering
		\caption{The performance of various methods to estimate \textbf{subgroup ATTs} in the simulations with the \textbf{misspecified PS model (PS2, see Section 3.1)}. The true treatment effect for subgroup 1 to 4 are $-10$, $-10/3$, $10/3$, and $10$, respectively. The number of Monte Carlo repetitions is 500. \label{tabs2}}%
		\begin{tabular*}{450pt}{@{\extracolsep\fill}ccccccccc@{\extracolsep\fill}}
			\toprule
			 & \textbf{Outcome}  & \textbf{Subgroup}  & \textbf{Logistic}  & \textbf{Logistic-S} & \textbf{CBPS} & \textbf{SBPS} & \textbf{G-SBPS} & \textbf{kG-SBPS} \\
			\midrule
			\multirow{8}{*}{\%Bias}& \multirow{4}{*}{Standard}& \textbf{1} & -241.55 & -34.98 & -299.36 & -34.98 & -3.40 & -13.31 \\&& \textbf{2} & -285.46 & -34.71 & -464.46 & -34.71 & 0.01 & -7.35 \\ && \textbf{3} & -267.60 & -80.45 & -112.08 & -80.17 & -0.15 & -11.35 \\ && \textbf{4} & -165.89 & -55.06 & -123.82 & -55.06 & -0.10 & -11.36 \\ 
            \cline{2-9}
            &\multirow{4}{*}{Extended}& \textbf{1} & 45.82 & 29.88 & 39.41 & 29.88 & 21.15 & 7.62 \\&& \textbf{2} & 2.08 & 96.21 & -79.89 & 96.21 & 105.05 & 6.36 \\ && \textbf{3} & -293.19 & -148.20 & -173.98 & -148.02 & -81.50 & -15.65 \\ && \textbf{4} & -206.03 & -84.18 & -162.27 & -84.18 & -18.31 & -17.92 \\ 
			\midrule
			\multirow{8}{*}{RMSE}& \multirow{4}{*}{Standard}& \textbf{1} & 24.27 & 3.81 & 30.06 & 3.81 & 0.50 & 1.79 \\&& \textbf{2} & 10.03 & 1.38 & 15.86 & 1.38 & 0.12 & 0.57 \\ && \textbf{3} & 9.16 & 2.80 & 4.36 & 2.80 & 0.10 & 0.60 \\ && \textbf{4} & 16.68 & 5.79 & 12.47 & 5.79 & 0.16 & 1.45 \\ 
            \cline{2-9}
            &\multirow{4}{*}{Extended}& \textbf{1} & 5.41 & 3.85 & 4.97 & 3.85 & 2.44 & 1.12 \\&& \textbf{2} & 2.45 & 3.50 & 3.75 & 3.50 & 3.74 & 0.46 \\ && \textbf{3} & 9.97 & 5.10 & 6.18 & 5.10 & 2.90 & 0.71 \\ && \textbf{4} & 20.71 & 8.83 & 16.33 & 8.83 & 2.10 & 2.18 \\ 	
			\bottomrule
		\end{tabular*}
	\end{table}
\end{center}
\clearpage

\begin{center}
	\begin{table}[t]%
            \renewcommand{\thetable}{S\arabic{table}}
            \footnotesize
		\centering
		\caption{The performance of various methods to estimate \textbf{subgroup ATTs} in the simulations with the \textbf{misspecified PS model (PS2, see Section 3.1)}. The simulation setting is the same as Table \ref{tabs2} except that the sample size per subgroup is doubled at $N_k = 1000$.  \label{tabs3}}%
		\begin{tabular*}{450pt}{@{\extracolsep\fill}ccccccccc@{\extracolsep\fill}}
			\toprule
			 & \textbf{Outcome}  & \textbf{Subgroup}  & \textbf{Logistic}  & \textbf{Logistic-S} & \textbf{CBPS} & \textbf{SBPS} & \textbf{G-SBPS} & \textbf{kG-SBPS} \\
			\midrule
			\multirow{8}{*}{\%Bias}& \multirow{4}{*}{Standard}& \textbf{1} & -241.72 & -33.73 & -299.05 & -33.73 & -11.76 & -8.71 \\&& \textbf{2} & -275.92 & -33.55 & -450.34 & -33.55 & -1.76 & -5.63 \\ && \textbf{3} & -257.87 & -78.71 & -106.73 & -78.48 & -0.31 & -6.72 \\ && \textbf{4} & -165.83 & -53.84 & -124.16 & -53.84 & 0.08 & -7.68 \\ 
            \cline{2-9}
            &\multirow{4}{*}{Extended}& \textbf{1} & 46.60 & 29.03 & 40.87 & 29.03 & 22.57 & 5.57 \\&& \textbf{2} & 11.55 & 99.93 & -67.43 & 99.93 & 108.39 & 6.97 \\ && \textbf{3} & -283.78 & -145.66 & -168.21 & -145.53 & -80.29 & -13.09 \\ && \textbf{4} & -205.95 & -83.08 & -162.64 & -83.08 & -17.65 & -13.55 \\ 
			\midrule
			\multirow{8}{*}{RMSE}& \multirow{4}{*}{Standard}& \textbf{1} & 24.23 & 3.54 & 29.97 & 3.54 & 1.32 & 1.05 \\&& \textbf{2} & 9.46 & 1.23 & 15.21 & 1.23 & 0.13 & 0.36 \\ && \textbf{3} & 8.74 & 2.68 & 3.94 & 2.68 & 0.08 & 0.35 \\ && \textbf{4} & 16.63 & 5.54 & 12.46 & 5.54 & 0.11 & 0.94 \\ 
            \cline{2-9}
            &\multirow{4}{*}{Extended}& \textbf{1} & 5.07 & 3.37 & 4.59 & 3.37 & 2.44 & 0.74 \\&& \textbf{2} & 1.87 & 3.49 & 2.97 & 3.49 & 3.75 & 0.38 \\ && \textbf{3} & 9.59 & 4.94 & 5.85 & 4.94 & 2.76 & 0.53 \\ && \textbf{4} & 20.65 & 8.55 & 16.32 & 8.55 & 1.94 & 1.56 \\ 	
			\bottomrule
		\end{tabular*}
	\end{table}
\end{center}
\clearpage

\begin{figure}[!htbp]
\centerline{\includegraphics[scale = 1]{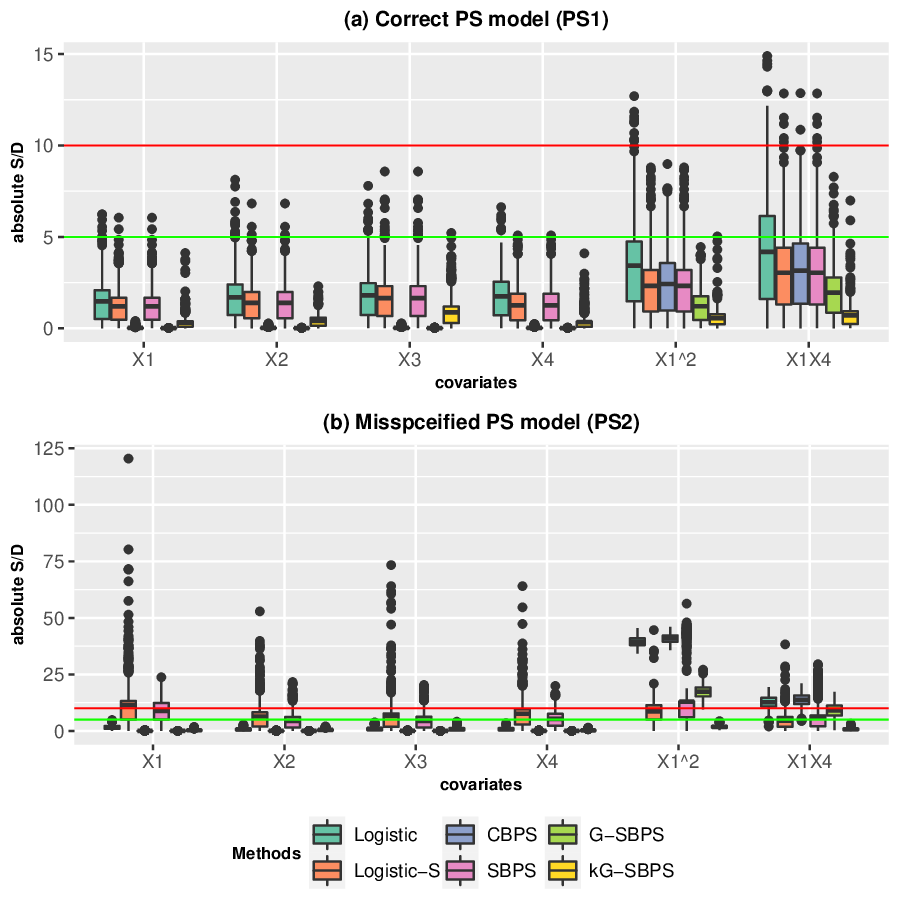}}
 \captionsetup{labelformat=empty}
 \caption{Figure S1. Boxplots of the standardized differences (S/D; \%) in the overall population when estimating the \textbf{ATE} in the simulation studies. The data are simulated from the correct PS model (PS1) or the misspecified PS model (PS2). The boxplots show the distribution of S/D from 500 Monte Carlo repetitions. The red and green horizontal lines mark the 10\% and 5\% S/D, respectively. } \label{FigS1}
\end{figure}
\clearpage

\begin{figure}[!htbp]
\centerline{\includegraphics[scale = 1]{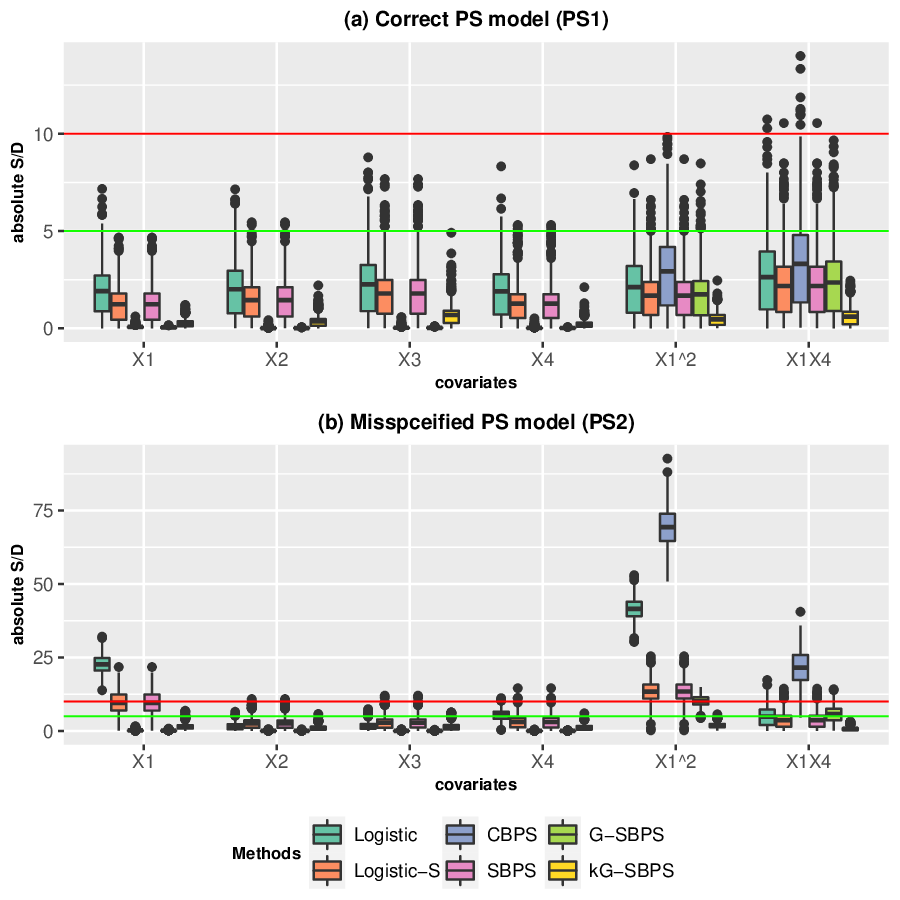}}
 \captionsetup{labelformat=empty}
 \caption{Figure S2. Boxplots of the standardized differences (S/D; \%) in the overall population when estimating the \textbf{ATT} in the simulation studies. The data are simulated from the correct PS model (PS1) or the misspecified PS model (PS2). The boxplots show the distribution of S/D from 500 Monte Carlo repetitions. The red and green horizontal lines mark the 10\% and 5\% S/D, respectively.}
 \label{FigS2}
\end{figure}

\begin{figure}[!htbp]
\centerline{\includegraphics[scale = 1]{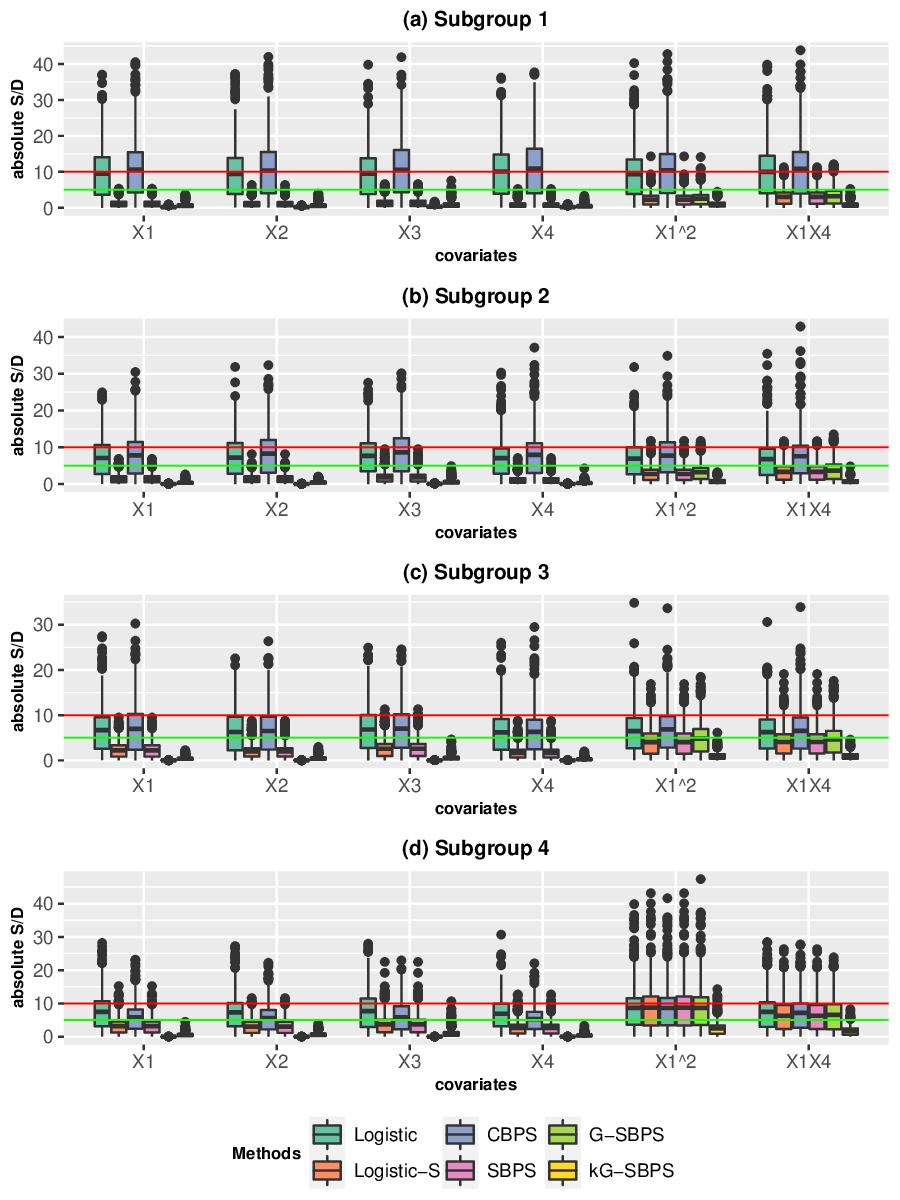}}
 \captionsetup{labelformat=empty}
 \caption{Figure S3. Boxplots of the subgroup S/D in the estimation of \textbf{ATT} from the simulations. The data are simulated from the correct PS model (PS1; see Section 3.1). The boxplots show the distribution of S/D from 500 Monte Carlo repetitions. The red and green horizontal lines mark the 10\% and 5\% S/D, respectively.}
   \label{FigS3}
\end{figure}

\begin{figure}[!htbp]
\centerline{\includegraphics[scale = 1]{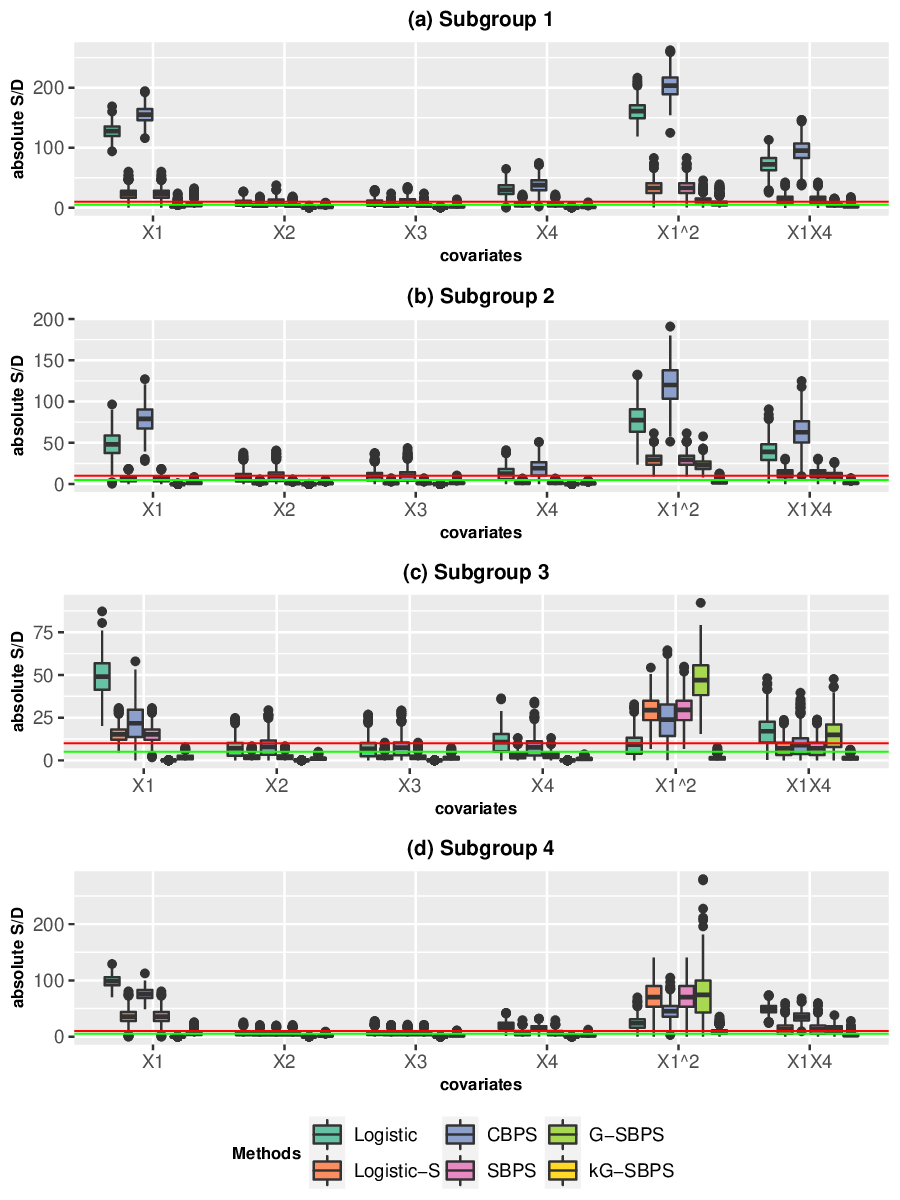}}
 \captionsetup{labelformat=empty}
 \caption{Figure S4. Boxplots of the subgroup S/D in the estimation of \textbf{ATT} from the simulations. The data are simulated from the misspecified PS model (PS2; see Section 3.1). The boxplots show the distribution of S/D from 500 Monte Carlo repetitions. The red and green horizontal lines mark the 10\% and 5\% S/D, respectively.}
   \label{FigS4}
\end{figure}

\begin{center}
\singlespacing
	\begin{table}[t]%
            \renewcommand{\thetable}{S\arabic{table}}
            \small
		\centering
		\caption{Description of the subgroups in the RHC data application. $N_0$: the number of non-RHC patients; $N_1$ the number of RHC patients.\label{tabs4}}%
		\begin{tabular*}{450pt}{@{\extracolsep\fill}clcc@{\extracolsep\fill}}
			\toprule
                &&\multicolumn{2}{c}{\textbf{Sample Size}} \\
			\textbf{Subgroup}  &  \textbf{Description} & $N_0$ & $N_1$ \\
			\midrule
			\multirow{4}{*}{3 subgroups (Non-overlapped)}& Mean Blood Pressure & & \\& subgroup 1:  $< 80$  & 2028& 1690\\
            & subgroup 2: $80 \ge$ and $\le 120$ & 679& 229\\
            & subgroup 3: $> 120$ & 844& 264\\
			\midrule
			\multirow{8}{*}{6 subgroups (Overlapped)}& Mean Blood Pressure & & \\ & subgroup 1: $< 80$  & 2028& 1690\\
            & subgroup 2: $80 \ge$ and $\le 120$ & 679& 229\\
            & subgroup 3: $> 120$ & 844& 264\\ & Estimated Probability of Surviving 2 Months & & \\ & subgroup 4: $<0.3$ & 302 & 238 \\ & subgroup 5: $\geq 0.3$ and $\leq 0.7$ & 1910 & 1296 \\ & subgroup 6: $> 0.7$ & 1339 & 649 \\ 
			\bottomrule
		\end{tabular*}
	\end{table}
\end{center}
\clearpage

\begin{center}
\singlespacing	
 \begin{table}[t]%
            \renewcommand{\thetable}{S\arabic{table}}
            \footnotesize
		\centering
		\caption{The estimated subgroup ATEs in the RHC data using the different propensity score analysis methods. The treatment effect measures the average increase in the hospital length of stay in days between RHC and non-RHC. \label{tabs5}}%
		\begin{tabular*}{470pt}{@{\extracolsep\fill}cccccccc@{\extracolsep\fill}}
			\toprule
			\textbf{Subgroup}  & & \textbf{Logistic} & \textbf{Logistic-S} & \textbf{CBPS} & \textbf{SBPS}& \textbf{G-SBPS} & \textbf{kG-SBPS}\\
			\midrule
			\multirow{3}{*}{3 subgroups (Non-overlapped)}& subgroup 1 & 1.77 & 1.98 & 1.26 & 1.98 & 1.69 & 1.75 \\& subgroup 2 & 5.65 & 1.97 & 5.95 & 1.97 & 1.61 & 1.13 \\
            & subgroup 3 & 4.88 & 3.47 & 4.89 & 3.47 & 2.34 & 3.38 \\
			\midrule
			\multirow{6}{*}{6 subgroups (Overlapped)}& subgroup 1 & 1.71 & - & 1.21 & - & 1.28 & 1.05 \\ & subgroup 2 & 5.46 & - & 5.58 & - & 0.84 & 1.48 \\
            & subgroup 3 & 4.92 & - & 4.84 & - & 2.70 & 3.10 \\ & subgroup 4 & 4.35 & - & 3.64 & - & 3.86 & 2.32 \\ & subgroup 5 & 2.51 & - & 2.05 & - & 0.58 & 1.82 \\
            & subgroup 6 & 3.03 & - & 3.12 & - & 2.34 & 0.87 \\
			\bottomrule
		\end{tabular*}
	\end{table}
\end{center}
\clearpage

\begin{figure}[!htbp]
\centerline{\includegraphics[scale = 1]{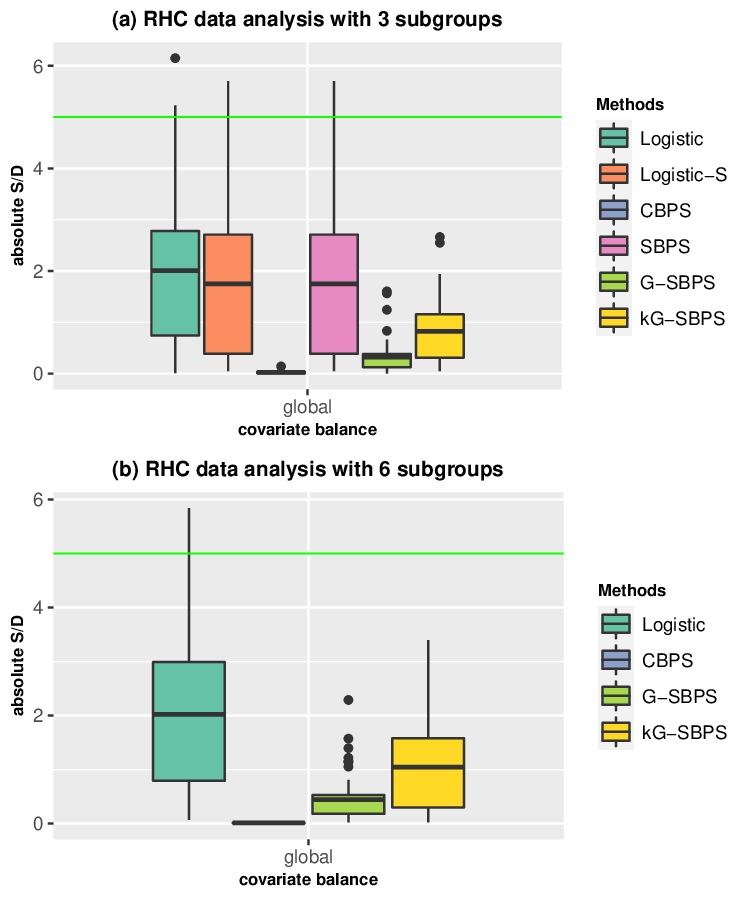}}
 \captionsetup{labelformat=empty}
 \caption{Figure S5. Boxplots of the standardized differences (S/D) of all covariates in the RHC data analysis. Green line: 5\% S/D. The S/D is calculated for the overall population (``global'' covariance balance). }
   \label{FigS5}
\end{figure}

\begin{center}
\singlespacing
	\begin{table}[t]%
            \renewcommand{\thetable}{S\arabic{table}}
            \footnotesize
		\centering
		\caption{Baseline covariates of subjects in the DSMT data. Mean (standard deviation) is reported for age and percent of residents living below poverty. Count (percentage) is reported for all other categorical variables. \label{tabs6}}%
		\begin{tabular*}{460pt}{@{\extracolsep\fill}lcc@{\extracolsep\fill}}
			\toprule
			\textbf{Covariate}  &  \textbf{Untreated} & \textbf{DSMT} \\
            & $n = 16401$ & $n = 3283$ \\
			\midrule
			 \textbf{Age (in years)}  & 75.3 (7.2) & 75.3 (6.3)\\
              \textbf{Percent of residents living below poverty (\%)} & 14.6 (9.3) & 15.2 (9.5) \\
              \textbf{Female sex} - n(\%) & 8048 (49.1) & 1611 (49.1) \\
              \textbf{Type II diabetes} - n(\%) & 15891 (96.9) & 3181 (96.9) \\
              \textbf{Type I diabetes} - n(\%) & 516 (3.1) & 106 (3.2) \\
              \textbf{Incident diabetes} - n(\%) & 2450 (14.9) & 490 (14.9) \\
              \textbf{Race} (Hispanic) - n(\%) & & \\ White & 11410 (69.6) & 2168 (66.0) \\
              Black & 1931 (11.8) & 375 (11.4) \\
              \textbf{Cancer category} (Colorectal cancer) - n(\%) & & \\ Prostate cancer & 6763 (41.2) & 1393 (42.4) \\
              Breast cancer & 6717 (41.0) & 1340 (40.8) \\
              \textbf{Married} - n(\%) & 6855 (41.8) & 1503 (45.8) \\ 
              \textbf{Medicare original entitlement status} - n(\%) & 1301 (7.9) & 277 (8.4) \\
              \textbf{Elixhauser comorbidity index category} ($0-1$) - n(\%) &  & \\
              $1-2$ & 5922 (36.1) & 1254 (38.2) \\
              $>3$ & 6723 (41.0) & 1250 (38.1)\\
              \textbf{Quantile of zip code percent non-high school graduates} (Very poor) - n(\%) & & \\
              Very good & 4202 (25.6) & 811 (24.7) \\
              Good & 3924 (23.9) & 707 (21.5) \\
              Poor & 3602 (22.0) & 731 (22.3) \\
              \textbf{Medicaid dual eligible during index year} - n(\%) & 2380 (14.5) & 589 (17.9)\\
              \textbf{Primary care provider in the year prior the index year} - n(\%) & 9202 (56.1) & 1601 (48.8) \\
              \textbf{Living in metropolitan} - n(\%) & 2959 (18.0) & 692 (21.1) \\ \textbf{Diabetes complication category} (No complication)- n(\%) & & \\ Renal manifestation & 1677 (10.2) & 386 (11.8) \\
              Ophthalmic manifestation & 2069 (12.6) & 567 (17.3) \\
              Neurological manifestation & 3618 (22.1) & 941 (28.7) \\
              Unspecified & 1094 (6.7) & 267 (8.1) \\
              \textbf{Elixhauser comorbidity category} - n(\%) & & \\
              Alcohol abuse & 195 (1.2) & 46 (1.4) \\
              Arrhythmia & 4289 (26.2) & 847 (25.8) \\
              Blood loss Anemia & 379 (2.3) & 70 (2.1) \\
              Congestive heart failure & 2964 (18.1) & 543 (16.5) \\
              COPD (chronic obstructive pulmonary disease) & 4092 (24.9) & 688 (21.0) \\
              Coagulopathy & 802 (4.9) & 158 (4.8) \\
              Deficiency anemia & 2570 (15.7) & 444 (13.5) \\
              Depression & 2639 (16.1) & 432 (13.2) \\
              Drug abuse & 224 (1.4) & 34 (1.0) \\
              Fluid and electrolyte disorders & 2801 (17.1) & 496 (15.1) \\
              AIDS/HIV & 12 (0.1) & NA* \\
              Hypothyroidism & 4612 (28.2) & 864 (26.3) \\
              Liver disease & 1224 (7.5) & 259 (7.9) \\
              Obesity & 2531 (15.4) & 557 (17.0) \\
              Other neurological disorders & 1403 (8.6) & 187 (5.7) \\
              Pulmonary circulation disorders & 674 (4.1) & 120 (3.7) \\
              Peptic ulcer disease excluding bleeding & 252 (1.5) & 50 (1.5) \\
              Peripheral vascular disorders & 3580 (21.8) & 626 (19.1) \\
              Paralysis & 275 (1.7) & 41 (1.2) \\
              Psychoses & 475 (2.9) & 61 (1.9) \\
              Renal failure & 3010 (18.4) & 680 (20.7)\\
              Rheumatoid arthritis & 1057 (6.4) & 187 (5.7) \\
              Valvular disease & 2513 (15.3) & 500 (15.2) \\
              Weight loss & 875 (5.3) & 122 (3.7) \\
              Hypertension & 14152 ( 86.3) & 2892 (88.1)\\
			\bottomrule
		\end{tabular*}
  \begin{tablenotes}
       \item [1] \small *The AIDS/HIV in DSMT group contains fewer than 11 patients. It was not reported because of the CMS cell size suppression policy to protect the confidentiality of entrollees.
     \end{tablenotes}
	\end{table}
\end{center}
\clearpage


\begin{center}
\singlespacing
	\begin{table}[t]%
            \renewcommand{\thetable}{S\arabic{table}}
            \footnotesize
		\centering
		\caption{Description of the subgroups in the DSMT data application. $N_0$: the number of DSMT patients; $N_1$ the number of non-DSMT patients. \label{tabs7}}%
		\begin{tabular*}{450pt}{@{\extracolsep\fill}clcc@{\extracolsep\fill}}
			\toprule
                &&\multicolumn{2}{c}{\textbf{Sample Size}} \\
			\textbf{Subgroup}  &  \textbf{Description} & $N_0$ & $N_1$ \\
			\midrule
			\multirow{5}{*}{4 subgroups (Non-overlapped)}& Living area and Medicaid dual eligibility during index year & & \\& subgroup 1:  Non-metropolitan and Non-eligible& 11599 & 2201\\
            & subgroup 2: Non-metropolitan and Eligible & 1843 & 390\\
            & subgroup 3: Metropolitan and Non-eligible & 2422 & 493\\ & subgroup 4: Metropolitan and Eligible & 537 & 199 \\
			\midrule
			\multirow{8}{*}{6 subgroups (Overlapped)}& Living area and Medicaid dual eligibility during index year  & & \\& subgroup 1:  Non-metropolitan and Non-eligible & 11599 & 2201\\
            & subgroup 2: Non-metropolitan and Eligible & 1843 & 390\\
            & subgroup 3: Metropolitan and Non-eligible & 2422 & 493\\ & subgroup 4: Metropolitan and Eligible & 537 & 199 \\ & Marriage status & & \\ & subgroup 5: Not married & 9546 & 1780 \\ & subgroup 6: Married & 6855 & 1503 \\  
			\bottomrule
		\end{tabular*}
	\end{table}
\end{center}
\clearpage

\begin{center}
\singlespacing	
 \begin{table}[t]%
            \renewcommand{\thetable}{S\arabic{table}}
            \footnotesize
		\centering
		\caption{The estimated subgroup ATEs in the DSMT data using the different propensity score analysis methods. The treatment effect measures the average increase in the hospitalization rate (\%) within 3 years of index year between DSMT and non-DSMT. \label{tabs8}}%
		\begin{tabular*}{470pt}{@{\extracolsep\fill}cccccccc@{\extracolsep\fill}}
			\toprule
			\textbf{Subgroup}  & & \textbf{Logistic} & \textbf{Logistic-S} & \textbf{CBPS} & \textbf{SBPS}& \textbf{G-SBPS} & \textbf{kG-SBPS}\\
			\midrule
			\multirow{4}{*}{4 subgroups (Non-overlapped)}& subgroup 1 & -3.8 & -4.0 & -3.8 & -4.0 & -4.0 & -3.8 \\& subgroup 2 & -12.6 & -6.4 & -12.6 & -6.4 & -6.6 & -5.4 \\
            & subgroup 3 & -4.3 & -8.8 & -4.3 & -8.8 & -9.1 & -9.4 \\ & subgroup 4 & -26.2 & -15.1 & -26.5 & -15.1 & -16.9 & -19.1 \\
			\midrule
			\multirow{6}{*}{6 subgroups (Overlapped)}& subgroup 1 & -3.8 & - & -3.7 & - & -3.9 & -3.9 \\ & subgroup 2 & -12.6 & - & -12.7 & - & -6.6 & -5.5 \\
            & subgroup 3 & -4.3 & - & -4.0 & - & -9.7 & -9.1 \\ & subgroup 4 & -26.2 & - & -27.2 & - & -17.6 & -19.7 \\ & subgroup 5 & -10.4 & - & -10.5 & - & -8.3 & -8.5 \\
            & subgroup 6 & -1.4 & - & -1.2 & - & -3.1 & -2.7 \\
			\bottomrule
		\end{tabular*}
	\end{table}
\end{center}
\clearpage

\begin{figure}[!htbp]
\centerline{\includegraphics[scale = 1]{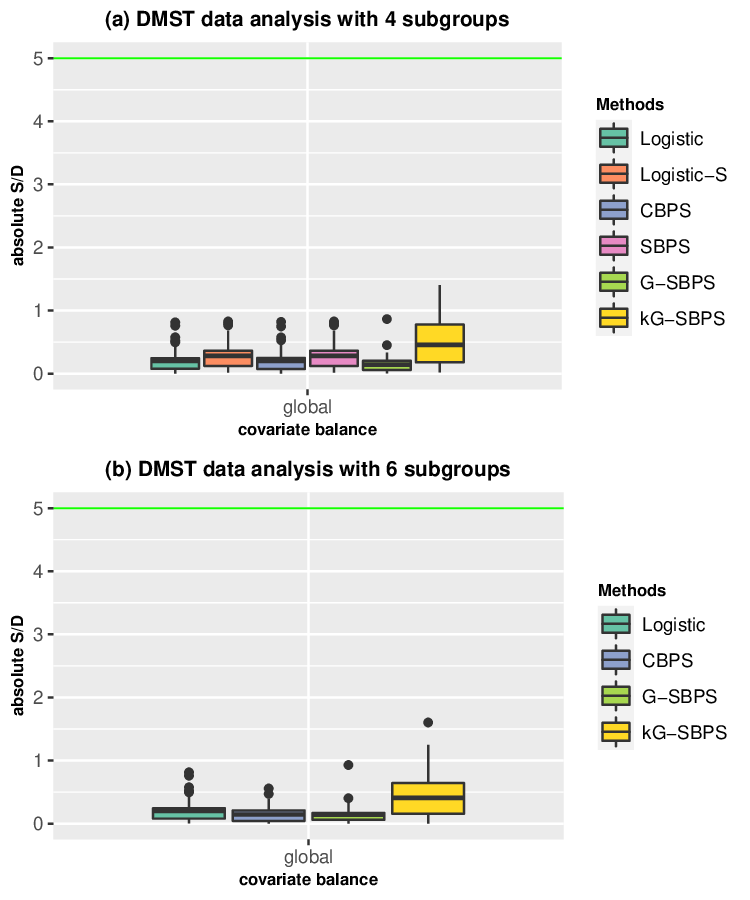}}
 \captionsetup{labelformat=empty}
 \caption{Figure S6. Boxplots of the standardized differences (S/D) of all covariates in the DSMT data analysis. Red line: 10\% S/D; Green line: 5\% S/D. The S/D is calculated for the overall population (``global'' covariance balance).}
   \label{FigS6}
\end{figure}

\begin{figure}[!htbp]
\centerline{\includegraphics[scale = 1]{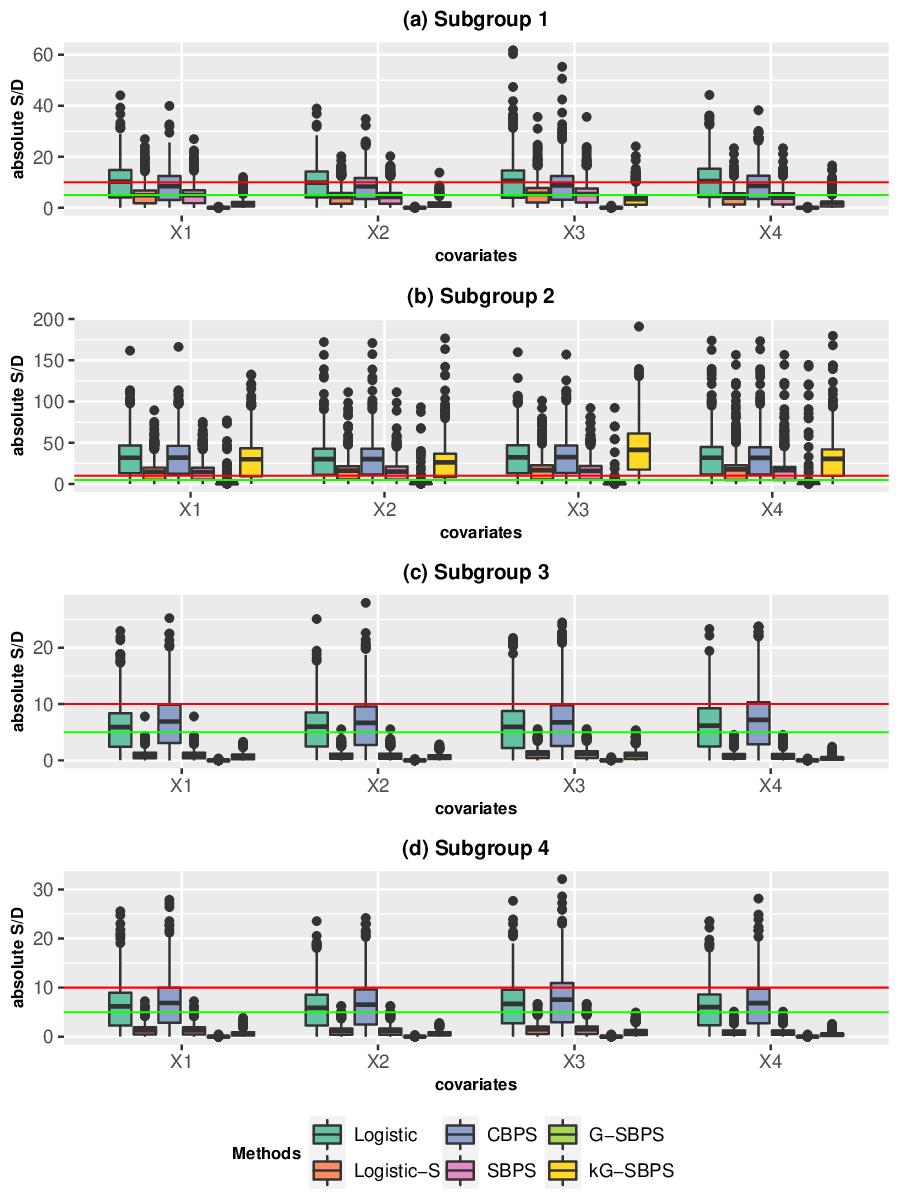}}
 \captionsetup{labelformat=empty}
 \caption{Figure S7. Boxplots of the S/D in four subgroups by different methods using IPW estimator to estimate \textbf{ATE}. The simulation scenario is Correct PS model (PS1; see Section 3.1), with 4 subgroups. $N_2 = 40$ for subgroup 2 and $N_k = 500$ for subgroup $k \in [1,3,4]^T$. The number of simulations is 500. Red line: 10\% S/D; Green line: 5\% S/D; S/D: standardized difference.}
   \label{FigS7}
\end{figure}

\begin{figure}[!htbp]
\centerline{\includegraphics[scale = 1]{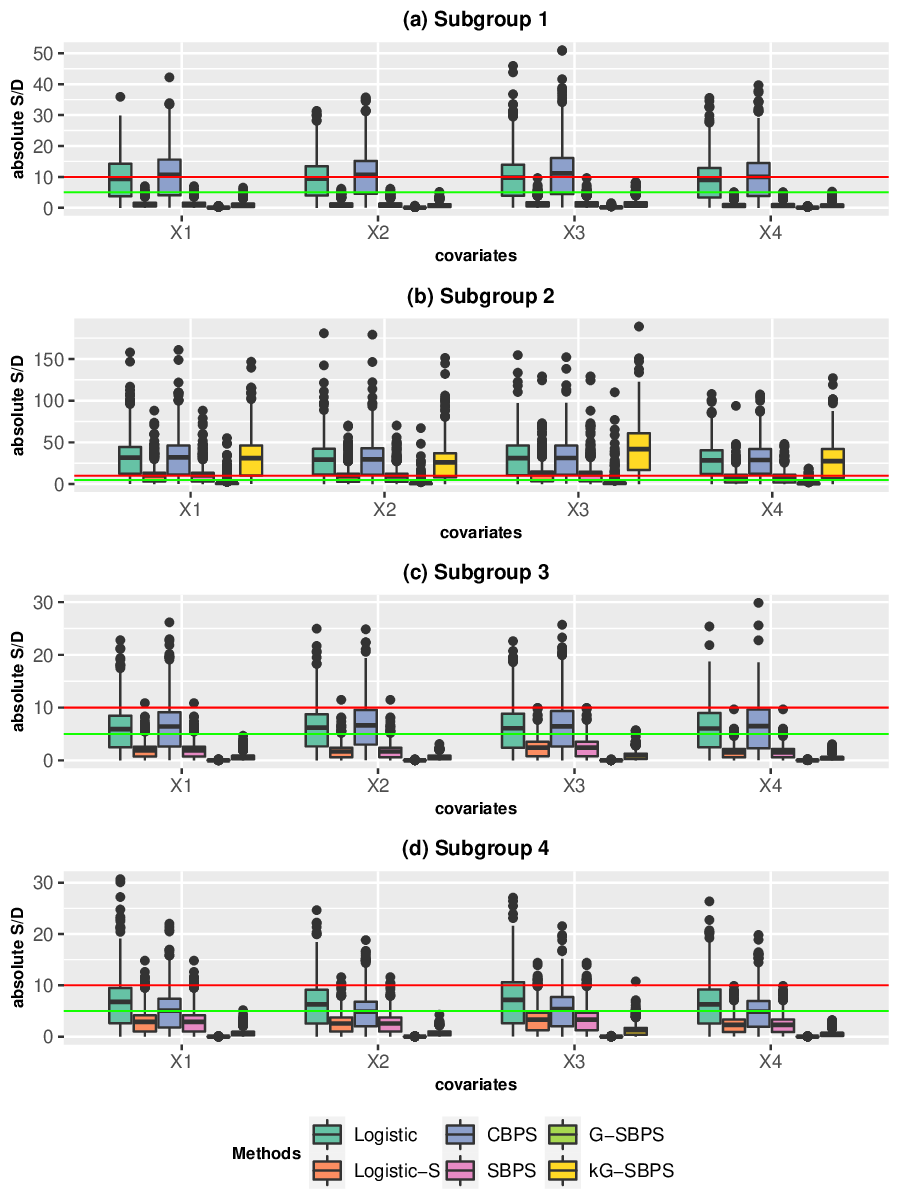}}
 \captionsetup{labelformat=empty}
 \caption{Figure S8. Boxplots of the S/D in four subgroups by different methods using IPW estimator to estimate \textbf{ATT}. The simulation scenario is Correct PS model (PS1; see Section 3.1), with 4 subgroups. $N_2 = 40$ for subgroup 2 and $N_k = 500$ for subgroup $k \in [1,3,4]^T$. The number of simulations is 500. Red line: 10\% S/D; Green line: 5\% S/D; S/D: standardized difference.}
   \label{FigS8}
\end{figure}

\end{document}